\documentclass[pra,aps,showpas,floatfix,superscriptaddress,twocolumn,letter]{revtex4}
\usepackage{graphicx}
\usepackage{bm} \usepackage{amssymb}
\usepackage{amsmath} \usepackage{amsfonts}
\usepackage{subfigure}

\newcommand{\ket}[1]{|#1\rangle}
\newcommand{\bra}[1]{\langle#1|}

\newcommand{\dtime}[1]{\frac{\partial#1}{\partial t}}

\renewcommand{\Re}{\text{Re}}
\renewcommand{\Im}{\text{Im}}

\begin{document}
\title{Lossless anomalous dispersion and an inversionless gain doublet via dressed interacting ground states (DIGS)}
\author{James Owen Weatherall}
\affiliation{Department of Physics and Engineering Physics,
Stevens Institute of Technology, Castle Point on Hudson, Hoboken, NJ 07030, USA}
\affiliation{Department of Mathematical Sciences,
Stevens Institute of Technology, Castle Point on Hudson, Hoboken, NJ 07030, USA}
\affiliation{Department of Logic and Philosophy of Science,
UC Irvine, 3151 Social Science Plaza A, Irvine, CA 92697, USA}
\author{Christopher P. Search}
\affiliation{Department of Physics and Engineering Physics,
Stevens Institute of Technology, Castle Point on Hudson, Hoboken, NJ 07030, USA}

\date{\today}

\begin{abstract}
Transparent media exhibiting anomalous dispersion have been of
considerable interest since Wang, Kuzmich, and Dogariu [Nature {\bf 406},
277 (2000)] first observed light propagate with superluminal and negative
group velocities without absorption. Here, we propose an atomic model
exhibiting these properties, based on a generalization of amplification
without inversion in a five-level DIGS system.  The system consists of a
$\Lambda$ atom prepared as in standard electromagnetically induced
transparency (EIT), with two additional metastable ground states coupled
to the $\Lambda$ atom ground states by two RF/microwave fields. We
consider two configurations by which population is incoherently pumped
into the ground states of the atom. Under appropriate circumstances, we
predict a pair of new gain lines with tunable width, separation, and
height.  Between these lines, absorption vanishes but dispersion is large
and anomalous.  The system described here is a significant improvement
over other proposals in the anomalous dispersion literature in that it
permits additional coherent control over the spectral properties of the
anomalous region, including a possible $10^4$-fold increase over the group
delay observed by Wang, Kuzmich, and Dogariu.
\end{abstract}
\maketitle

\section{Introduction}

It has long been predicted that under certain conditions, optical media
exhibit anomalous dispersion (ie. the index of refraction increases with
increasing wavelength) \cite{Rayleigh}.  In these cases, the group
velocity of a pulse of light with appropriate frequency can be larger than
the speed of light in a vacuum or negative \cite{Brillouin}.  Negative
group velocity implies that a smoothly varying pulse of light will appear
to exit the dispersive medium before it enters \cite{Garrett+McCumber}.
This is possible because different spectral components of the pulse
interfere strongly in the dispersive region and transform the leading edge
of the pulse into an image of the pulse's peak.  Thus it is only with
regard to the overall profile of the pulse that superluminal propagation
appears to occur; anomalous dispersion does not permit superluminal
signalling and causality is not violated \cite{Diener, Macke+Segard,
WKD-PRL, Huang+Zhang}.

In 2000, Wang, Kuzmich, and Dogariu were first to observe light
propagating with negative group velocities in a region of low absorption
and minimal reshaping in a Rb vapor cell \cite{WKD-Nature, WKD-PRA}. Their
experiment was based on a system proposed by Steinberg and Chiao
\cite{Steinberg+Chiao}, in which a $\Lambda$ atom with the excited state
coupled to one ground state by two far-detuned pump lasers is probed by a
weak beam near the transition between the excited state and the second
ground state. With appropriate initial conditions, this configuration
leads to two narrowly spaced Raman gain lines in the probe's spectrum,
with a region of anomalous dispersion but low absorption between them.  In
2003, Bigelow et al. reported superluminal (and subluminal) group
velocities in a room temperature solid (specifically, an alexandrite
crystal) \cite{Bigelow+etal1, Bigelow+etal2}. That same year, the claim
that causality is maintained in cases of superluminal group velocities was
supported experimentally by Stenner et al., who showed that the detection
of a non-analytic point in an incident wave (representing new information)
on the far side of a region of anomalous dispersion took at least as long
as in the vacuum case \cite{Stenner+etal}.  It is now well established
that superluminal and negative group velocities are possible in otherwise
transparent and non-interfering media without violating relativity or
causality.

In the decade since these first experiments, systems exhibiting anomalous
dispersion have received considerable attention in the literature, among
theorists and experimentalists.  Some of these studies followed up on
early predictions that strongly-driven two level atoms
\cite{Wilson-Gordon+Friedmann,  Szymanowski+Keitel, Quang+Freedhoff,
Wicht+etal, Rocco+etal} and degenerate three level atoms
\cite{Friedmann+Wilson-Gordon, Rosenhouse-Danstker+etal} can also exhibit
regions of large anomalous dispersion but low absorption (for a
comparative study of systems of these sorts and additional references, see
\cite{Wicht+etal-Comp}).  Others have involved coupling the excited states
of a $\mathtt{V}$ atom \cite{Bortman-Arbiv+etal} or the ground states of a
$\Lambda$ atom \cite{Agarwal+etal} and then using the coupling field to
coherently control the sign of the dispersion, while others still have
employed a magnetic field to induce a Zeeman splitting, to similar effect
\cite{Ghulghazaryan+Malakyan, Goren+etal}.   Many of these subsequent
studies have involved coherent modification of the Raman gain process
proposed by Steinberg and Chiao by introducing a second excited state
\cite{Kang+etal1, Kang+etal2, Agarwal+Dasgupta, Hu+etal, Chen+etal,
Tajalli+Sahrai, Sahrai+etal} or a second excited state and a third ground
state \cite{Wang+etal}.

In the current contribution, we present a novel proposal for producing a
narrow, closely-spaced gain doublet with an intermediate window of
anomalous dispersion, using dressed interacting ground states (DIGS). DIGS
systems \cite{Weatherall+Search, Weatherall+etal} are a generalization of
the double dark resonances (sometimes called interacting dark resonances)
introduced by Lukin et al. \cite{Lukin+etal}.  Double dark resonances
occur when one ground state of a $\Lambda$ atom is strongly coupled to the
excited state (as in standard EIT), and also coupled to a third ground
state via an RF/microwave field.  The result is a standard EIT absorption
spectrum, with a new absorption peak located at zero probe detuning.  New
absorption nulls appear to either side of this peak.  These systems have
been studied extensively and observed experimentally \cite{Lukin+etal,
Yan+etal, Yelin+etal, Ye+etal, Mahmoudi+etal, Chen+etal-IDR, Wilson+etal}.
DIGS systems include a fourth ground state (to produce a five-level atom)
coupled to the second $\Lambda$ ground state by a second RF/microwave
field.  This additional coupling has been predicted to split the double
dark resonance absorption peak into two symmetric peaks located within the
transparency window \cite{Weatherall+Search}. These peaks have widths and
locations that are tunable by varying the RF/microwave Rabi frequencies,
leading to additional control over the optical response of an atomic
system.

Here, we modify the model presented in \cite{Weatherall+Search} by
introducing pumping terms.  The pumping transforms the lines we describe
in the earlier paper into the pair of gain lines.  The system studied here
has several benefits over previously considered examples exhibiting
anomalous dispersion.  For one, even in the cases were the Raman gain
lines were shown to have tunable heights and signs, there was no
independent control over their widths.  Moreover, here the locations of
the gain lines vary with the field strengths, not frequency or phase, and
so they are easier to tune than a Raman system, where the location of the
peaks depends on the laser frequencies.  The lines we predict have tunable
locations, widths, and heights (allowing the smooth change of the sign of
both the absorption coefficient and dispersion), and so the current system
permits a very narrow window with ultrahigh anomalous dispersion, or
alternatively, a maximally broad window throughout which anomalous
dispersion is present. We predict that in a cold Rb gas prepared in
analogy to the experiment in Ref. \cite{Kash+etal}, one could observe a
negative group velocity index two orders of magnitude larger than the
largest yet observed of $-14000$ \cite{Kim+etal}, and four orders of
magnitude larger than the group velocity index reported by Wang, Kuzmich,
and Dogariu \cite{WKD-Nature}.

Since the gain described here is generated without appreciable population
accumulating in the excited state, the phenomenon we predict can be
thought of as a generalized example of amplification without inversion
(AWI), which is often also referred to as lasing without inversion
\cite{Kocharovskaya, Mompart+Corbalan}. Unlike standard AWI \cite{Scully,
Kocharovskaya+Khanin, Harris-LWI}, however, the current system requires
only incoherent pumping to exhibit gain, as opposed to coherent pumping.
We derive analytic results to describe the optical response as a function
of the incoherent pumping rate(s) in both open and closed pumping
configurations.  These analytic results are a central feature of the
current paper, as previous presentations of related systems (eg. double
dark state systems with pumping) present exclusively numerical solutions.

The remainder of the paper will be organized as follows. In section
\ref{model}, we will review the base model that we consider in this paper.
For further details on the model, see \cite{Weatherall+Search}.  In
section \ref{solutions} we will solve the master equations for both open
and closed pumping configurations.  In section \ref{susceptibility}, we
will derive the linear susceptibility for the probe beam.  Here, we will
present an explanation and analysis of the gain lines and the anomalous
dispersion mentioned above.  Section \ref{doppler} will treat Doppler
broadening in the DIGS system.  Finally, in section \ref{conclusion} we
will offer some conclusions, including a discussion of possible
experimental realizations.

\section{Model Preliminaries}\label{model}

\begin{figure}\centering
\includegraphics[width=.5\columnwidth,angle=270,trim= 2in 1in 2in 1in,clip]{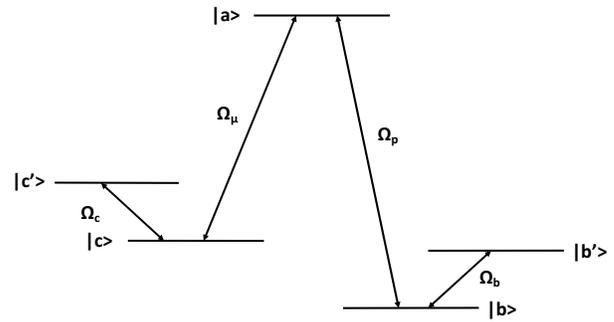}
\caption{\label{fig-Model}Our five level model.  An excited state $\ket{a}$ is coupled to two lower energy state doublets, $\{\ket{b},\ket{b'}\}$ and $\{\ket{c},\ket{c'}\}$.  $\Omega_{\mu}$ is the Rabi frequency of a strong control beam coupling $\ket{a}$ and $\ket{c}$; $\Omega_b$ and $\Omega_c$ are the Rabi frequencies of two RF/microwave fields coupling the members of each of the doublets.  We study the propagation of a weak probe beam with Rabi frequency $\Omega_p\ll\Omega_c,\Omega_b,\Omega_{\mu}$ near resonance with the $\ket{a}\leftrightarrow\ket{b}$ transition.  The specifics of the decay and pumping schemes will be treated in sections \ref{open} and \ref{closed}.}
\end{figure}

Our base system is a five-level atom (see Fig. \ref{fig-Model}) in which two sets of ground states, $\{\ket{b},\ket{b'}\}$ and $\{\ket{c},\ket{c'}\}$, interact with a single excited state, $\ket{a}$.  Transitions between each of the ground state doublets are assumed to be dipole forbidden; the members of the hyperfine ground state doublets, meanwhile, are coupled via RF/microwave fields (in what follows, we will refer to these as RF fields for simplicity) of frequencies $\omega_b$ and $\omega_c$.  These fields have Rabi frequencies $\Omega_b$ and $\Omega_c$.  One of the states in the $\{\ket{c},\ket{c'}\}$ manifold (without loss of generality, $\ket{c}$) is strongly coupled to $\ket{a}$ by a field with Rabi frequency $\Omega_{\mu}$.  We study the optical response of a field near resonance with the $\ket{b}\leftrightarrow\ket{a}$ transition, with Rabi frequency $\Omega_p$.  We describe this system via the Hamiltonian
\begin{align}
\tilde{\mathcal{H}}=&\frac{\hbar}{2}\left(\omega_a\ket{a}\bra{a}+(\omega_b+\nu_p)\ket{b}\bra{b}+(\omega_{b'}+\nu_p-\nu_b)\ket{b'}\bra{b'}\right.\notag\\
&+(\omega_c+\nu_{\mu})\ket{c}\bra{c}+(\omega_{c'}+\nu_{\mu}-\nu_c)\ket{c'}\bra{c'}-\Omega_{\mu} \ket{a}\bra{c}\notag\\
&-\Omega_{b}\ket{b'}\bra{b}
-\Omega_{c}\ket{c'}\bra{c}
\left.-\Omega_{p}\ket{a}\bra{b}\right)+\text{h.c.}
\end{align}
The $\tilde{}$ here indicates that we have written the Hamiltonian in a
rotating basis defined to eliminate explicit time dependence. From the
Hamiltonian we obtain the detunings for the control laser
$\Delta_\mu=\omega_a-\omega_c-\nu_{\mu}$ and the RF fields,
$\Delta_b=\omega_{b'}-\omega_b-\nu_b$ and
$\Delta_c=\omega_{c'}-\omega_c-\nu_c$. Most importantly, the detuning of
the probe laser from the $\ket{b}\leftrightarrow \ket{a}$ transition
\begin{equation}
\Delta_p=\omega_a-\omega_b-\nu_p
\end{equation}
will be used to express the optical susceptibility of the probe field.

For further details on this base system (including an explicit definition
of the rotating frame), see section II of \cite{Weatherall+Search}. The
model considered there functions as the starting point for the current
paper; it is essentially identical to the model described here, except for
the pumping and decay schemes, representations of which will be discussed
presently.

In what follows, we will be interested in the linear susceptibility
expanded around the $\ket{a}\leftrightarrow\ket{b}$ transition, which can
be written in terms of the corresponding density matrix element,
$\tilde{\rho}_{ab}$. Since we are not yet considering any nonunitary
contributions, the density matrix can be described by the von Neumann
equation, $i\hbar d\rho/dt=[H,\rho]$. The equations of motion for the
components of the density matrix follow immediately.  They are given in
Appendix A. In what follows, we will make reference to these in defining
the models for each of the pumping configurations.

In the bare basis, we will find that to first order in
$\Omega_p/\Omega_{\mu}$ (with assumptions that we describe below)
$\tilde{\rho}_{ab}$ depends on the solutions to six linked differential
equations, which in general cannot be solved analytically.  However, under
the rotation defined in \cite{Weatherall+Search}, in which we diagonalize
the $\{\ket{b},\ket{b'}\}$ and $\{\ket{c},\ket{c'}\}$ subspaces of the
Hamiltonian, these six linked equations decouple into two systems of three
equations each.  The details of the change of basis are given in
appendices B and C.  We will distinguished the diagonalized states by
using capital letters,
$\{\ket{b},\ket{b'}\}\rightarrow\{\ket{B},\ket{B'}\}$ and
$\{\ket{c},\ket{c'}\}\rightarrow\{\ket{C},\ket{C'}\}$. The transformation
for each of the subspaces individually is of the same form as the dressed
states of a two level atom.

\section{Solutions to the master equation}\label{solutions}

Here we present and solve the Linblad equation for the evolution of the
density matrix in both open (section \ref{open}) and closed (section
\ref{closed}) pumping configurations.  In each case, we are focused on
finding a linear solution for $\tilde{\rho}_{ab}$ from which we can derive
the linear susceptibility for the probe beam in section
\ref{susceptibility}.

\subsection{Open pumping configuration}\label{open}

As mentioned in the introduction, the amplification presented here does
not require population inversion.  It does require, however, that
population be distributed between the ground states--specifically in
$\{\ket{b},\ket{b'}\}$ and$\ket{c'}$.  The details of this requirement are
most clearly manifest when we populate these states directly, via pumping
from unspecified external states.  This is the case we consider in the
current section: the ground states $\ket{b}$ and $\ket{c'}$ are pumped
from external levels; likewise, decay occurs to external levels. Note that
we would obtain essentially the same results if we added additional
pumping to the levels $\ket{b'}$ and $\ket{c}$ since $\Omega_b$
distributes the population between $\ket{b}$ and $\ket{b'}$ while in the
limit we are interested in ($\Omega_c \ll \Omega_{\mu}$), any population
in $\ket{c}$ is optically pumped into $\ket{c'}$.

To model pumping from and decay to external states, we modify Eqs.
\ref{baseEquations} by including terms to model direct incoherent pumping
to $\ket{b}$ and $\ket{c'}$, at rates $r_b$ and $r_{c'}$ respectively.
\begin{align*}
&i\dtime{\tilde{\rho}_{jj}}\sim ir_j& &j=b,c'&
\end{align*}
Relaxation terms are modeled by,
\begin{align*}
&i\dtime{\tilde{\rho}_{jj}}\sim-i\gamma_j\rho_{jj}&\\
&i\dtime{\tilde{\rho}_{jk}}\sim-i\gamma_{jk}\rho_{jk}& &j\neq k&
\end{align*}
where $\gamma_j$ is the decay from state $\ket{j}$ and $\gamma_{jk}=\frac{1}{2}(\gamma_j+\gamma_k)+\gamma_{jk}^{\text{ph}}$ is the full off-diagonal relaxation term.

Our strategy in what follows will be to assume that
$\Omega_p\ll\Omega_b,\Omega_c,r_b,r_{c'}\ll\Omega_{\mu}$, and then use two
applications of perturbation theory.  First, we work at zeroth order in
$\Omega_p/\Omega_{\mu}$, which essentially decouples the
$\{\ket{b},\ket{b'}\}$ manifold from $\ket{a}$, $\ket{c}$, and $\ket{c'}$.
The $\{\ket{b},\ket{b'}\}$ subspace can be solved exactly at this order.
The $\{\ket{a},\ket{c},\ket{c'}\}$ subspace is more complicated.  However,
under the approximations already described $\tilde{\rho}_{c'c'}$ varies
slowly relative to the other terms of the density matrix in this subspace,
and so we can assume the other terms will follow it adiabatically.  Then
we can solve for the other terms to first order in $\Omega_c/\Omega_{\mu}$
as a function of the steady state population of $\ket{c'}$,
$\tilde{\rho}^{\text{st}}_{c'c'}$,  and use these first order solutions to
find a self-consistent solution for $\tilde{\rho}^{\text{st}}_{c'c'}$
valid to order $(\Omega_c/\Omega_{\mu})^2$. Finally, we will move to the
dressed basis introduced in section \ref{model} and defined in appendix B
and solve for $\tilde{\rho}_{ab}$ to first order in
$\Omega_p/\Omega_{\mu}$ using the zeroth order (in
$\Omega_p/\Omega_{\mu}$) solutions as source terms.

We assume that the control and RF fields are on resonance,
$\Delta_{\mu}=\Delta_b=\Delta_c=0$. (A full general solution for the
susceptibility with arbitrary nonzero detunings is included in appendix D.
The derivation is identical to the current case.)  Then, to zeroth order
in $\Omega_p/\Omega_{\mu}$, the equations of motion for the
$\{\ket{b},\ket{b'}\}$ manifold can be written as
\begin{subequations}
\begin{align}
i\dtime{\tilde{\rho}_{bb}}&= ir_b-i\gamma_b\tilde{\rho}_{bb}+\frac{\Omega_{b}}{2}(\tilde{\rho}_{bb'}-\tilde{\rho}_{b'b})\\
i\dtime{\tilde{\rho}_{b'b'}}&= -i\gamma_{b'}\tilde{\rho}_{b'b'}-\frac{\Omega_{b}}{2}(\tilde{\rho}_{bb'}-\tilde{\rho}_{b'b})\\
i\dtime{\tilde{\rho}_{bb'}}&= -\gamma_{bb'}\tilde{\rho}_{bb'} +\frac{\Omega_{b}}{2}(\tilde{\rho}_{bb}-\tilde{\rho}_{b'b'})
\end{align}
\end{subequations}
These can be solved by writing them in the form $\dtime{X}=-M X+A$, which has a steady state solution of $X=M^{-1} A$.  We find
\begin{subequations}
\begin{align}
\tilde{\rho}_{bb}^{\text{st}}&=\frac{r_b(2\gamma_{b'}\gamma_{bb'}+\Omega_b^2)}{2\gamma_{b}\gamma_{b'}\gamma_{bb'} +(\gamma_{b}+\gamma_{b'})\Omega_b^2}\\
\tilde{\rho}_{b'b'}^{\text{st}}&=\frac{r_b\Omega_b^2}{2\gamma_{b}\gamma_{b'}\gamma_{bb'} +(\gamma_b+\gamma_{b'})\Omega_b^2}\\
\tilde{\rho}_{bb'}^{\text{st}}=(\tilde{\rho}_{b'b}^{\text{st}})^*&=\frac{-ir_b\gamma_{b'}\Omega_b}{2\gamma_b\gamma_{b'}\gamma_{bb'} +(\gamma_b+\gamma_{b'})\Omega_b^2}.
\end{align}
\end{subequations}

To zeroth order in $\Omega_p/\Omega_{\mu}$, our equations of motion for the $\{\ket{a},\ket{c},\ket{c'}\}$ subspace are, first for the diagonal terms,
\begin{subequations}
\begin{align}
i\dtime{\rho_{aa}}&=-i\gamma_a\rho_{aa}-\frac{\Omega_{\mu}}{2}(\rho_{ca}-\rho_{ac})\\
i\dtime{\rho_{cc}}&=-i\gamma_c\rho_{cc}+\frac{\Omega_{\mu}}{2}(\rho_{ca}-\rho_{ac})+\frac{\Omega_c}{2}(\rho_{cc'}-\rho_{c'c})\\
i\dtime{\rho_{c'c'}}&=i r_{c'}-i\gamma_{c'}\rho_{c'c'}-\frac{\Omega_c}{2}(\rho_{cc'}-\rho_{c'c})
\end{align}
and for the off-diagonal terms,
\begin{align}
i\dtime{\tilde{\rho}_{ca}}&= -i\gamma_{ca}\tilde{\rho}_{ca} +\frac{\Omega_{\mu}}{2}(\tilde{\rho}_{cc} -\tilde{\rho}_{aa}) -\frac{\Omega_{c}}{2}\tilde{\rho}_{c'a}\\
i\dtime{\tilde{\rho}_{c'a}}&= -i\gamma_{c'a}\tilde{\rho}_{c'a} +\frac{\Omega_{\mu}}{2}\tilde{\rho}_{c'c} - \frac{\Omega_{c}}{2}\tilde{\rho}_{ca}\\
i\dtime{\tilde{\rho}_{c'c}}&= -i\gamma_{c'c}\tilde{\rho}_{c'c} +\frac{\Omega_{\mu}}{2}\tilde{\rho}_{c'a} +\frac{\Omega_{c}}{2}(\tilde{\rho}_{c'c'} - \tilde{\rho}_{cc})
\end{align}
\end{subequations}

We assume that
$\Omega_{\mu},\gamma_a\gg\Omega_c,r_{c'},\gamma_c,\gamma_{c'}$.  With
these assumptions, it is clear that all of the equations of motion, except
the one governing $\tilde{\rho}_{c'c'}$, are dominated by the terms
proportional to $\Omega_{\mu}$ and $\gamma_a$.  This is our justification
for the claim that $\tilde{\rho}_{c'c'}$ varies slowly relative to the
other terms in this subspace, and that therefore $\tilde{\rho}_{c'c'}$ can
be treated as a constant with respect to the other equations of motion.

To zeroth order in $\Omega_c/\Omega_{\mu}$, we have two decoupled systems
of homogeneous equations. The steady state occurs only when
$\tilde{\rho}^{(0)}_{c'a}=\tilde{\rho}^{(0)}_{c'c}=\tilde{\rho}^{(0)}_{aa}=\tilde{\rho}^{(0)}_{cc}=\tilde{\rho}^{(0)}_{ca}=0$.
Physically, this makes sense, since the system is non-conservative and
there is no external pumping to these levels/coherences.  To first order
in $\Omega_c/\Omega_{\mu}$,
$\tilde{\rho}^{(1)}_{aa}=\tilde{\rho}^{(1)}_{cc}=\tilde{\rho}^{(1)}_{ca}=0$
again, since the equations of motion are unchanged at this order.  The
second system, however, now leads to nonzero steady state values.  These
are
\begin{align}
\tilde{\rho}^{(1)}_{c'a}&=\tilde{\rho}^{\text{st}}_{c'c'}\frac{-\Omega_c\Omega_{\mu}}{4\gamma_{c'c}\gamma_{c'a}+\Omega_{\mu}^2}\\
\tilde{\rho}^{(1)}_{c'c}&=\tilde{\rho}^{\text{st}}_{c'c'} \frac{-2i\gamma_{c'a}\Omega_c}{4\gamma_{c'c}\gamma_{c'a}+\Omega_{\mu}^2}
\end{align}
This linear solution is sufficient to reproduce the effect that we are
interested in. Taking the first order solutions as the steady states, we
can solve self-consistently for $\tilde{\rho}_{c'c'}^{\text{st}}$.  We
find
\begin{equation}
\tilde{\rho}_{c'c'}^{\text{st}}=\frac{r_{c'}}{\left(\frac{2\Omega_c^2\gamma_{c'a}}{4\gamma_{c'c}\gamma_{c'a}+\Omega_{\mu}^2} +\gamma_{c'}\right)}.
\end{equation}

We are now very nearly in a position to solve for $\tilde{\rho}_{ab}$ to
first order in $\Omega_p/\Omega_{\mu}$. We move now to the dressed basis
defined in appendix B.  The assumptions already stated,
$\Delta_b=\Delta_c=0$, imply that $\Omega_b^{\text{eff}}=\Omega_b$,
$\Omega_c^{\text{eff}}=\Omega_c$, and $\theta_b=\theta_c=\pi/4$ (see
appendix B for definitions of these terms). In order to handle decay
analytically, we assume that $\gamma_{ab}=\gamma_{ab'}$.  This is
reasonable, supposing both expressions will be dominated by $\gamma_a \gg
\gamma_b,\gamma_b',\gamma^{ph}_{a,b},\gamma^{ph}_{a,b'}$. Moreover, we
assume that $\gamma_b\approx\gamma_{b'}$,
$\gamma^{\text{ph}}_{bc}\approx\gamma^{\text{ph}}_{b'c}$, and
$\gamma^{\text{ph}}_{bc'}\approx\gamma^{\text{ph}}_{b'c'}$ so that we can
take $\gamma_{cb}\approx\gamma_{cb'}=\gamma_C$ and
$\gamma_{c'b}\approx\gamma_{c'b'}=\gamma_{C'}$. In Appendix C we give full
expressions for the decay and dephasing of the relevant density matrix
components in the dressed basis in terms of the $\gamma_{ab}$,
$\gamma_{C}$, and $\gamma_{C'}$. These approximations may seem slightly
arbitrary, but they permit both enough simplification to solve the problem
entirely, and yet contain enough nuance for an adequate analysis of the
effects of decoherence on the phenomena we predict. Finally, we assume
that $\gamma_{cc'},\gamma_{bb'},\gamma_C,\gamma_{C'}\ll\Omega_{\mu}$, as
would occur, say, in a cold atomic gas, or in a hot gas with a buffer gas
present.

We already have that $\tilde{\rho}^{(1)}_{aa}=0$; moreover, under these
new assumptions, $\tilde{\rho}^{(1)}_{c'a}=
-\frac{\Omega_c}{\Omega_{\mu}}\tilde{\rho}^{\text{st}}_{c'c'}$, which
gives that (see appendix B)
$\tilde{\rho}_{C'a}=\tilde{\rho}_{Ca}=-\frac{\sqrt{2}}{2}\frac{\Omega_c}{\Omega_{\mu}}\tilde{\rho}^{\text{st}}_{c'c'}$.
Meanwhile,
$\tilde{\rho}_{bb}\approx\tilde{\rho}_{b'b'}\approx\frac{r_b}{\gamma_b+\gamma_{b'}}$
and
$\tilde{\rho}_{bb'}\approx\frac{-ir_b\gamma_{b'}}{(\gamma_b+\gamma_{b'})\Omega_b}$.
The diagonalization leaves these invariant, and so
$\tilde{\rho}_{BB}\approx\tilde{\rho}_{B'B'}\approx
\frac{r_b}{\gamma_b+\gamma_{b'}}$ and
$\tilde{\rho}_{BB'}=(\tilde{\rho}_{B'B})^*\approx\frac{-ir_b\gamma_{b'}}{(\gamma_b+\gamma_{b'})\Omega_b}$.
Taking these together, we find steady state solutions
\begin{align}\label{simpRhoaB}
\tilde{\rho}_{aB}&=\frac{\sqrt{2}\Omega _p }{2Z_+}\left( \mathfrak{P}_B \left(2 i\gamma _{C}-2 \Delta _p-\Omega _b\right)\right.\notag\\
&\;\;\left.\times \left(2i \gamma _{C'}-2 \Delta _p-\Omega _b\right)+\Omega_c^2\left(\tilde{\rho}_{c'c'}^{\text{st}}- \mathfrak{P}_B\right) \right)
\end{align}
and
\begin{align}\label{simpRhoaB'}
\tilde{\rho}_{aB'}&=-\frac{\sqrt{2}\Omega _p}{2Z_-}\left(\mathfrak{P}_B \left(2 i\gamma _{C}-2\Delta _p+\Omega _b\right)\right.\notag\\
&\;\;\left.\times \left(2i \gamma _{C'}-2 \Delta _p+\Omega _b\right)+\Omega_c^2\left(\tilde{\rho}^{\text{st}}_{c'c'}- \mathfrak{P}_B\right) \right)
\end{align}
where
\begin{align}\label{simpZ}
Z_{\pm}&=\Omega_{\mu}^2(2i\gamma_{C'}-2\Delta_p\mp\Omega_b)-\left(2i \gamma_{ab}-2 \Delta _p\mp\Omega _b\right)\notag\\
&\;\;\times\left(\left(i\gamma _{C}+i\gamma _{C'}-2 \Delta _p\mp\Omega _b \right)^2-\Omega _c^2\right)
\end{align}
and where we have defined
$\mathfrak{P}_B=\frac{r_b(\Omega_b-i\gamma_{b'})}{(\gamma_b+\gamma_{b'})\Omega_b}$
(for a general definition of $\mathfrak{P}_B$, see appendix D). Meanwhile,
to be consistent with the other approximations made thus far we should
note that $\tilde{\rho}^{\text{st}}_{c'c'}$ simplifies to
\begin{equation}
\tilde{\rho}_{c'c'}^{\text{st}}=\frac{r_{c'}\Omega_{\mu}^2}{2\gamma_{c'a}\Omega_c^2+\gamma_{c'}\Omega_{\mu}^2}.
\end{equation}
$\tilde{\rho}_{ab}$ can be found simply from Eqs. \ref{simpRhoaB} and
\ref{simpRhoaB'} via the relation
$\tilde{\rho}_{ab}=\frac{\sqrt{2}}{2}\left(\tilde{\rho}_{aB}-\tilde{\rho}_{aB'}\right)$.

\subsection{Closed pumping configuration}\label{closed}

We chose to present the open pumping configuration first because we feel
it distills the important parts of the dynamics: as we will argue in
section \ref{susceptibility}, population in two ground states, $\ket{b}$
and $\ket{c'}$ is sufficient to produce amplification of the probe beam.
Thus the essential physics of the system is already present in Eqs.
\ref{simpRhoaB}-\ref{simpZ}; some readers may prefer to skip directly to
section \ref{susceptibility} now.  However, the theoretical literature has
tended to focus on closed systems.  Moreover, in the open pumping
configuration we pump $\ket{c'}$ directly, which begs the question of
whether population will accumulate in $\ket{c'}$ in the steady state if it
is not directly pumped there.  For completeness, we will now present a
more theoretically natural case, in which atoms are pumped directly from
$\ket{b}$ to $\ket{a}$, from which they decay to $\ket{b}$, $\ket{c}$, and
$\ket{c'}$.  In the appropriate limit ($\Omega_b/\Delta_b \rightarrow 0$),
the solution presented here is an analytic solution for the system
described in, for instance, \cite{Mahmoudi+etal}.

In the closed pumping case, atoms are pumped from $\ket{b}$ to $\ket{a}$,
and decay is internal to our 5 level subspace. We assume that the ground
states are stable for the purposes of the current calculation.  $\ket{a}$
is assumed to decay only to $\ket{b}$, $\ket{c}$, and $\ket{c'}$, with
branching ratios $\alpha_b$, $\alpha_c$, and $\alpha_{c'}$, respectively
($\alpha_b+\alpha_c+\alpha_{c'}=1$). Since a nonzero value of
$\tilde{\rho}_{c'c'}^{\text{st}}$ is necessary, it is crucial that
$\alpha_{c'}\neq 0$.  The base equations of motion, Eqs.
\ref{baseEquations}, now have contributions
\begin{subequations}
\begin{align}
i\dtime{\tilde{\rho}_{aa}}&\sim -i(\gamma_{a}+r)\tilde{\rho}_{aa}+ir\tilde{\rho}_{bb}\\
i\dtime{\tilde{\rho}_{bb}}&\sim i(\alpha_b\gamma_a+r)\tilde{\rho}_{aa}-ir\tilde{\rho}_{bb}\\
i\dtime{\tilde{\rho}_{cc}}&\sim i\alpha_c\gamma_a\tilde{\rho}_{aa}\\
i\dtime{\tilde{\rho}_{c'c'}}&\sim i\alpha_{c'}\gamma_a\tilde{\rho}_{aa}.
\end{align}
\end{subequations}
The definitions of the off-diagonal relaxation rates are unchanged from
the open pumping case, except that now
$\gamma_b=\gamma_{b'}=\gamma_c=\gamma_{c'}=0$.

Our strategy here will be the same as in the open pumping case.  To zeroth order in $\Omega_p/\Omega_{\mu}$, however, the two subsystems of the previous case no longer decouple.  But we can again make an observation about the time scales in the problem that will permit some simplification.  As before, we assume that $\Omega_{\mu}\gg\Omega_b,\Omega_c\gg r$; moreover, we take $\gamma_a$ to be sufficiently less that $\Omega_{\mu}$ for it to be the case that $\alpha_i\gamma_a$ is about an order of magnitude smaller than $\Omega_{\mu}$.  These considerations allow us to assume that $\tilde{\rho}_{bb}$, $\tilde{\rho}_{b'b'}$, $\tilde{\rho}_{bb'}$, and $\tilde{\rho}_{c'c'}$ vary slowly relative to the other elements.  We can thus assume that the rapidly varying ones follow these adiabatically.  We again work perturbatively in $\Omega_p/\Omega_{\mu}$, and solve for $\tilde{\rho}_{ac'}$ and $\tilde{\rho}_{c'c}$.  Now, however, $\tilde{\rho}_{bb}$ and $\tilde{\rho}_{c'c'}$ depend on $\tilde{\rho}_{aa}$, and so to find a self-consistent second order solution for the slowly varying populations, we require a second order perturbative solution for the rapidly varying populations.

We again work under the assumption that the control and RF field detunings vanish.  First, note that it is clear from inspection of the relevant equations of motion in Eqs. \ref{baseEquations} that in the steady state, $\tilde{\rho}^{\text{st}}_{bb}=\tilde{\rho}^{\text{st}}_{b'b'}$.  Meanwhile, we can solve for the rapidly varying terms perturbatively in $\Omega_c/\Omega_{\mu}$.  To zeroth order, we again find two sets of decoupled equations.  The ones describing $\tilde{\rho}_{ac'}$ and $\tilde{\rho}_{cc'}$ are homogeneous and decoupled from the pumping at this order, and so they vanish.  The second system now has an inhomogeneous term, $\tilde{\rho}_{bb}^{\text{st}}$,  We find,
\begin{subequations}
\begin{align}
i\dtime{\tilde{\rho}_{aa}}&=-i(\gamma_{a}+r)\tilde{\rho}_{aa} +ir\tilde{\rho}^{\text{st}}_{bb}
 -\frac{\Omega_{\mu}}{2}(\tilde{\rho}_{ca}-\tilde{\rho}_{ac})\\
i\dtime{\tilde{\rho}_{cc}}&= i\alpha_c\gamma_a\tilde{\rho}_{aa}  +\frac{\Omega_{\mu}}{2}( \tilde{\rho}_{ca} -\tilde{\rho}_{ac} )\\
i\dtime{\tilde{\rho}_{ca}}&= -i\gamma_{ca}\tilde{\rho}_{ca} +\frac{\Omega_{\mu}}{2}(\tilde{\rho}_{cc} -\tilde{\rho}_{aa} )
\end{align}
\end{subequations}
These have a steady state solution of
\begin{subequations}
\begin{align}
\tilde{\rho}^{(0)}_{aa}&=
\tilde{\rho}_{bb}\left(\frac{r}{r+(1-\alpha_c)\gamma_a}\right)\\
\tilde{\rho}^{(0)}_{cc}&=\tilde{\rho}_{bb}\left(\frac{r(2\alpha_c\gamma_a\gamma_{ca}^2+\gamma_{ca}\Omega_{\mu}^2)}{(r+(1-\alpha_c)\gamma_a)\gamma_{ca}\Omega_{\mu}^2}\right)\\
\tilde{\rho}^{(0)}_{ca}&=\tilde{\rho}_{bb}\left(\frac{r\alpha_c\gamma_a-i\gamma_{ca}}{(r+(1-\alpha_c)\gamma_a)\gamma_{ca}\Omega_{\mu}}\right)
\end{align}
\end{subequations}

To first order, the equations of motion for $\tilde{\rho}_{aa}$, $\tilde{\rho}_{cc}$, and $\tilde{\rho}_{ca}$ are unchanged, as the density matrix elements proportional to $\Omega_c$ are zero to zeroth order.  The equations for $\tilde{\rho}_{c'a}$ and $\tilde{\rho}_{c'c}$, meanwhile, become
\begin{subequations}
\begin{align}
i\dtime{\tilde{\rho}_{c'a}}&= -i\gamma_{c'a}\tilde{\rho}_{c'a} +\frac{\Omega_{\mu}}{2}\tilde{\rho}_{c'c} - \frac{\Omega_{c}}{2}\tilde{\rho}^{(0)}_{ca}\\
i\dtime{\tilde{\rho}_{c'c}}&= -i\gamma_{c'c}\tilde{\rho}_{c'c} +\frac{\Omega_{\mu}}{2}\tilde{\rho}_{c'a} +\frac{\Omega_{c}}{2}(\tilde{\rho}^{\text{st}}_{c'c'} - \tilde{\rho}^{(0)}_{cc})
\end{align}
\end{subequations}
These have a steady state solution of
\begin{subequations}
\begin{align}
\tilde{\rho}^{(1)}_{c'a}&=\frac{\Omega_c(\Omega_{\mu}(\tilde{\rho}^{(0)}_{cc}-\tilde{\rho}^{\text{st}}_{c'c'})+2i\gamma_{c'c}\tilde{\rho}^{(0)}_{ca})}{4\gamma_{c'c} \gamma_{c'a}+\Omega_{\mu}^2}\\
\tilde{\rho}^{(1)}_{c'c}&=\frac{\Omega_c(2i\gamma_{c'a}(\tilde{\rho}^{(0)}_{cc}-\tilde{\rho}^{\text{st}}_{c'c'})+\Omega_{\mu}\tilde{\rho}^{(0)}_{ca})}{4\gamma_{c'c}\gamma_{c'a} +\Omega_{\mu}^2}
\end{align}
\end{subequations}

In the open pumping case, it was only necessary to solve for the coherences to first order in $\Omega_c/\Omega_{\mu}$ in order to find the second order population $\tilde{\rho}_{c'c'}$.  In contrast, to find a fully self-consistent second order solution for $\tilde{\rho}_{c'c'}$ and $\tilde{\rho}_{bb}$ in the closed pumping case, we require a second order solution for $\tilde{\rho}_{aa}$ and $\tilde{\rho}_{cc}$.   The equations of motion for $\tilde{\rho}_{c'a}$ and $\tilde{\rho}_{c'c}$ are unchanged at this order.  The equations of motion for $\tilde{\rho}_{aa}$, $\tilde{\rho}_{cc}$, and $\tilde{\rho}_{ca}$ meanwhile are now
\begin{subequations}
\begin{align}
i\dtime{\tilde{\rho}_{aa}}&=-i(\gamma_{a}+r)\tilde{\rho}_{aa} +ir\tilde{\rho}_{bb}^{\text{st}}
 -\frac{1}{2}\Omega_{\mu}(\tilde{\rho}_{ca}-\tilde{\rho}_{ac})\\
i\dtime{\tilde{\rho}_{cc}}&= i\alpha_c\gamma_a\tilde{\rho}_{aa} -\frac{1}{2}\Omega_{c}(\tilde{\rho}^{(1)}_{c'c} -\tilde{\rho}^{(1)}_{cc'}) +\frac{1}{2}\Omega_{\mu}( \tilde{\rho}_{ca} -\tilde{\rho}_{ac} )\\
i\dtime{\tilde{\rho}_{ca}}&= -i\gamma_{ca}\tilde{\rho}_{ca} +\frac{1}{2}\Omega_{\mu}(\tilde{\rho}_{cc} -\tilde{\rho}_{aa} ) -\frac{1}{2}\Omega_{c}\tilde{\rho}^{(1)}_{c'a}
\end{align}
\end{subequations}
These are solved by
\begin{subequations}
\begin{align}
\tilde{\rho}^{(2)}_{aa}&=\frac{2r\tilde{\rho}^{\text{st}}_{bb}+i(\tilde{\rho}_{c'c}^{(1)}-\tilde{\rho}_{cc'}^{(1)})\Omega_c}{2(r+(1-\alpha_c)\gamma_a)}\\
\tilde{\rho}^{(2)}_{cc}&=\frac{1}{2(r+(1-\alpha_c)\gamma_a)\Omega_{\mu}^2}\times\left(r(4\alpha_c\gamma_a\gamma_{ca}+2\Omega_{\mu}^2)\tilde{\rho}_{bb}^{\text{st}}\right.\notag\\
&\;\;\left.+\,i\Omega_c\left(2(r+\gamma_a)\gamma_{ca}+\Omega_{\mu}^2\right)(\tilde{\rho}^{(1)}_{c'c}-\tilde{\rho}^{(1)}_{cc'})\right.\notag\\
&\;\;\left.+\,\Omega_c\Omega_{\mu}(r+(1-\alpha_c)\gamma_a)(\tilde{\rho}_{ac'}+\tilde{\rho}_{c'a})\right)\\
\tilde{\rho}^{(2)}_{ca}&=\frac{-2ir\alpha_c\gamma_a\tilde{\rho}_{bb}^{\text{st}} +(r+\gamma_a)(\tilde{\rho}_{c'c}^{(1)}-\tilde{\rho}_{cc'}^{(1)})\Omega_c}{2(r+(1-\alpha_c)\gamma_a)\Omega_{\mu}}\notag\\
&\;\;-\frac{i(\tilde{\rho}_{ac'}^{(1)}-\tilde{\rho}_{c'a}^{(1)})\Omega_c}{4\gamma_{ca}}.
\end{align}
\end{subequations}

The next step is to solve for $\tilde{\rho}_{bb}$ and
$\tilde{\rho}_{c'c'}$ self-consistently, in terms of these steady state
solutions. But first we can simplify the expressions already stated using
our initial approximations that $r\ll\Omega_c,\Omega_b\ll\Omega_{\mu}$.
Moreover, since the ground states are assumed not to decay,
$\gamma_{c'c}=\gamma_{c'c}^{\text{ph}}$.  We assume that dephasing effects
are small, and take $\gamma_{c'c}\ll r$.  Note that although we could in
principle proceed through the next step of the calculation without making
these assumptions, the expressions thus derived are unwieldy.  Moreover,
these assumptions will be necessary presently when we move to solve to
first order in $\Omega_p/\Omega_{\mu}$, and so it is expedient (and
consistent, given our initial assumptions) to make them now. We can write,
\begin{subequations}
\begin{align}
\tilde{\rho}^{(2)}_{c'a}&=-\left(\frac{\Omega_c}{\Omega_{\mu}}\right)\tilde{\rho}_{c'c'}\\
\tilde{\rho}^{(1)}_{c'c}&=-\left(\frac{2i\gamma_{c'a}\Omega_c}{\Omega_{\mu}^2}\right)\tilde{\rho}_{c'c'}\\
\tilde{\rho}^{(2)}_{aa}&=\frac{r\Omega^2_{\mu}\tilde{\rho}_{bb}+2\gamma_{c'a}\Omega_c^2\tilde{\rho}_{c'c'}}{(1-\alpha_c)\gamma_a\Omega_{\mu}^2} \\
\tilde{\rho}^{(2)}_{cc}&=\frac{1}{(1-\alpha_c)\gamma_a\Omega_{\mu}^2}\times\left(r\tilde{\rho}_{bb}(\alpha_c\gamma_a\gamma_{ca}+\Omega_{\mu}^2)\right.\notag\\
&\;\;\left.+2\frac{\Omega_c^2}{\Omega_{\mu}^2}(2\gamma_{c'a}\Omega_{\mu}^2+\gamma_a(4\gamma_{c'a}\gamma_{ca} -(1-\alpha_c)\Omega_{\mu}^2))\right)\\
\tilde{\rho}^{(2)}_{ca}& =\frac{-i(2\gamma_{c'a}\tilde{\rho}_{c'c'}\Omega_c^2+r\alpha_c\tilde{\rho}_{bb}\Omega_{\mu}^2)}{(1-\alpha_c)\Omega_{\mu}^3}
\end{align}
\end{subequations}

The equations of motion for $\tilde{\rho}_{bb}$ and $\tilde{\rho}_{c'c'}$ can now be written as
\begin{subequations}
\begin{align}
i\dtime{\tilde{\rho}_{bb}}&= i(\alpha_b\gamma_a+r)\tilde{\rho}^{(2)}_{aa}-ir\tilde{\rho}_{bb} \\
i\dtime{\tilde{\rho}_{c'c'}}&= i\alpha_{c'}\gamma_a\tilde{\rho}_{aa}^{(2)}+  \frac{\Omega_{c}}{2}(\tilde{\rho}^{(2)}_{c'c}-\tilde{\rho}^{(2)}_{cc'})
\end{align}
\end{subequations}
Inserting the expressions for the second order rapidly varying terms and combining the two resulting equations, we find the condition that the steady state populations $\tilde{\rho}_{bb}$ and $\tilde{\rho}_{c'c'}$ must satisfy.
\begin{equation}
\tilde{\rho}_{bb}=\frac{2\gamma_{c'a}\alpha_b}{r\alpha_{c'}}\left(\frac{\Omega_c^2}{\Omega_{\mu}^2}\right)\tilde{\rho}_{c'c'}
\end{equation}
This condition specifies a unique pair of populations when we impose the additional constraint that the sum of the populations must be 1.  Then,
\begin{align}\label{closedRhoc'c'}
\tilde{\rho}_{c'c'}=\frac{r\alpha_c\Omega_{\mu}^2}{4\alpha_b\gamma_{c'a}\Omega_c^2+r\alpha_c\Omega_{\mu}^2}.
\end{align}
The other populations, meanwhile, can now be written as
\begin{subequations}
\begin{align}
\tilde{\rho}_{aa}&\approx  0\\
\tilde{\rho}_{bb}=\tilde{\rho}_{b'b'}&\approx \frac{2\alpha_b\gamma_{c'a}\Omega_c^2}{4\alpha_b\gamma_{c'a}\Omega_c^2+r\alpha_c\Omega_{\mu}^2}\label{closedRhobb}\\
\tilde{\rho}_{cc}&\approx 0
\end{align}
\end{subequations}
$\tilde{\rho}_{aa}$ and $\tilde{\rho}_{cc}$ vanish because they are of
order
$\Omega_c^2r/(\Omega_{\mu}^2\gamma_a)\lesssim(\Omega_c/\Omega_{\mu})^3\ll
1$.

From here, the strategy is the same as in the open pumping case, and in
fact, the solution in that case carries over wholesale. Eqs.
\ref{simpRhoaB}, \ref{simpRhoaB'}, and \ref{simpZ} are general statements
in terms of the steady state solutions for $\tilde{\rho}_{aC}$,
$\tilde{\rho}_{aC'}$, and the $\tilde{\rho}_{aB}$, $\tilde{\rho}_{aB'}$
terms. In the semi-dressed basis introduced in section \ref{model}, we now
have $\tilde{\rho}_{aC}=\tilde{\rho}_{aC'}
=-\frac{\sqrt{2}}{2}\frac{\Omega_c}{\Omega_{\mu}}\tilde{\rho}_{c'c'}$,
just as in the open pumping case; meanwhile
$\tilde{\rho}_{BB}=\tilde{\rho}_{B'B'}=\tilde{\rho}_{bb}$, since
$\tilde{\rho}_{bb}=\tilde{\rho}_{b'b'}$ and $\tilde{\rho}_{bb'}=0$.  So if
we take $\mathfrak{P}_B=\tilde{\rho}_{BB}=
\frac{2\alpha_b\gamma_{c'a}\Omega_c^2}{4\alpha_b\gamma_{c'a}\Omega_c^2+r\alpha_c\Omega_{\mu}^2}$
and consider the closed pumping solution for $\tilde{\rho}_{c'c'}$, then
Eqs. \ref{simpRhoaB}-\ref{simpZ} hold, now as a function of the pumping
from $\ket{b}$ to $\ket{a}$, $r$.

\section{Linear response of the pumped DIGS systems}\label{susceptibility}

The complex linear susceptibility, expanded about the $\ket{a}\leftrightarrow\ket{b}$ transition, is given by
\begin{equation}
\chi^{(1)}=\frac{2 \sigma(\mathbf{r}) N D_{ab}}{\epsilon_0\mathcal{E}_p}\tilde{\rho}_{ab}.
\end{equation}
$\chi^{(1)}$ determines both the absorption coefficient,
$\alpha(\Delta_p)=k_p\Im[\chi^{(1)}(\Delta_p)]$, and the index of
refraction, $n(\Delta_p)\approx(1+\Re[\chi^{(1)}(\Delta_p)])^{1/2}$.
Considerations arising from the particular experimental set-up would
determine density profile, $\sigma(\mathbf{r})$, and number density of
atoms, $N$, which are included to formally account for the particular
optical thickness of the sample.  $D_{ab}=e\bra{a}\vec{\epsilon}\cdot
x\ket{b}$ is the dipole moment between $\ket{a}$ and $\ket{b}$, as a
function of laser polarization, $\vec{\epsilon}$. In our analysis we will
focus on the dimensionless reduced susceptibility
\begin{equation}\label{reduced_susceptibility}
\tilde{\chi}^{(1)}=\frac{\epsilon_0\hbar\gamma_{ab}}{D_{ab}^2N\sigma(\mathbf{r})}\chi^{(1)}=\frac{2\gamma_{ab}}{\Omega_p}\tilde{\rho}_{ab}.
\end{equation}
where we have used the definition of the probe Rabi frequency,
$\Omega_p=D_{ab}\mathcal{E}_p/\hbar$. This allows us to write the
following expression for the linear susceptibility as a function of
$\tilde{\rho}_{c'c'}$ and $\mathfrak{P}_B$, which will vary depending on
the pumping configuration.
\begin{widetext}
\begin{align}\label{fullSusceptibility}
\tilde{\chi}^{(1)}&=\gamma_{ab}\left(\frac{\left( \mathfrak{P}_B \left(2 i\gamma _{C}-2 \Delta _p-\Omega _b\right) \left(2i \gamma _{C'}-2 \Delta _p-\Omega _b\right)+\Omega_c^2\left(\tilde{\rho}_{c'c'}^{\text{st}}- \mathfrak{P}_B\right) \right)}{\Omega_{\mu}^2(2i\gamma_{C'}-2\Delta_p-\Omega_b)-\left(2i \gamma_{ab}-2 \Delta _p-\Omega _b\right)
\left(\left(i\gamma _{C}+i\gamma _{C'}-2 \Delta _p-\Omega _b \right)^2-\Omega _c^2\right)}\right.\notag\\
&\;\;+\left.\frac{\left(\mathfrak{P}_B \left(2 i\gamma _{C}-2\Delta _p+\Omega _b\right) \left(2i \gamma _{C'}-2 \Delta _p+\Omega _b\right)+\Omega_c^2\left(\tilde{\rho}^{\text{st}}_{c'c'}- \mathfrak{P}_B\right) \right)}{\Omega_{\mu}^2(2i\gamma_{C'}-2\Delta_p+\Omega_b)-\left(2i \gamma_{ab}-2 \Delta _p+\Omega _b\right)\left(\left(i\gamma _{C}+i\gamma _{C'}-2 \Delta _p+\Omega _b \right)^2-\Omega _c^2\right)}\right)
\end{align}
\end{widetext}

As a check on this solution, note that in the limit that
$\tilde{\rho}_{c'c'}\rightarrow 0$ (corresponding to $r_{c'}\rightarrow
0$) and $\mathfrak{P}_B\rightarrow 1/2$ (corresponding to normalized
population beginning in $\ket{b}$ and $\ket{b'}$), we recover the solution
presented in \cite{Weatherall+Search}, provided that
$\gamma_C,\gamma_{C'}$ are small; this latter solution, meanwhile, reduces
to the standard EIT solution in the limit that
$\Omega_b/\Delta_b,\Omega_c/\Delta_c\rightarrow 0$. Note that it also
amounts to an analytic solution to the double dark resonance system,
\cite{Lukin+etal}, in the limit that $\Omega_b/\Delta_b\rightarrow 0$.
(See Appendix D for $\tilde{\chi}^{(1)}$ for arbitrary $\Delta_{\mu}$,
$\Delta_b$, and $\Delta_c$)

In section \ref{gain}, we will examine the imaginary part of Eq. \ref{fullSusceptibility}, showing how the system(s) solved above lead to gain lines in the appropriate limits; section \ref{dispersion} will treat the real part of Eq. \ref{fullSusceptibility}, including the anomalous dispersion present between the gain lines.  This discussion requires us to estimate the values for the important variables in the problem. The Rabi frequencies of the coupling laser and RF fields are experimentally tunable over a large range. For the spontaneous emission rate, we take $\gamma_a=10^7 s^{-1}$ and will measure the Rabi frequencies and detunings in units of $\gamma_{ab}$. For the ground state dephasing rates the range of values are limited primarily by collision rates and therefore are temperature and density dependent. However for concreteness, we assume that $\tilde{\gamma}_{ab}$, $\gamma_C$, and $\gamma_{C'}$ are in the range $10^{3}-10^{4} s^{-1}$.

\subsection{Im$(\chi)$: Gain lines}\label{gain}

\begin{figure}
\includegraphics[width=1.0\columnwidth]{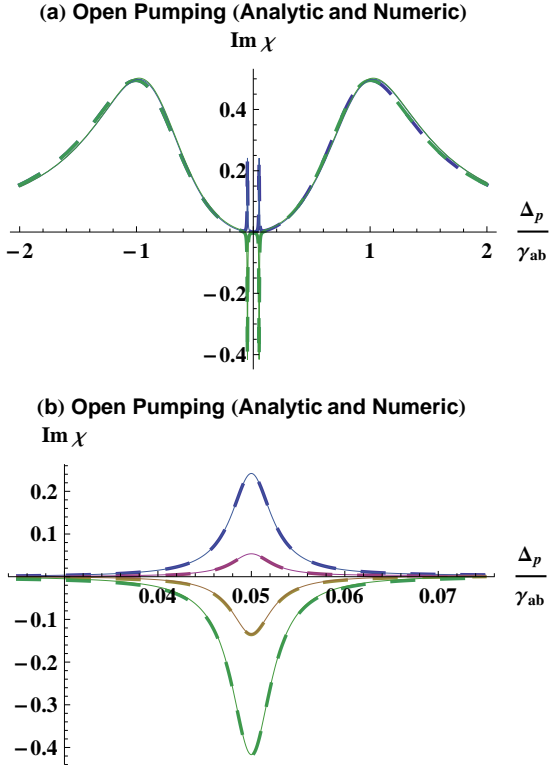}
\caption{\label{fig-openImaginaryParts} (Color online) Here we compare
analytic and numeric solutions for the imaginary part of the reduced
susceptibility (corresponding to the absorption coefficient) in the open
pumping case.  The numeric solutions are represented by wide dashed lines,
and the analytics by narrower solid lines.  Agreement between them is
quite good.  In the top plot, we show the full spectrum in the cases where
$r_{c'}=0$ (absorption lines; blue) and $r_{c'}=.007\gamma_{ab}$ (gain
lines; green).  In the bottom plot, we show a close up of one of the
narrow features.  The pumping parameters here, going from top to bottom,
are $r_{c'}=0$ (blue lines), $r_{c'}=.002\gamma_{ab}$ (red lines),
$r_{c'}=.004\gamma_{ab}$ (brown lines), and $r_{c'}=.007\gamma_{ab}$
(green lines).  In all cases, $r_b=.0001\gamma_{ab}$,
$\Omega_b=\Omega_c=.1\gamma_{ab}$, $\Omega_{\mu}=2\gamma_{ab}$, and
$\gamma_b=\gamma_{b'}=\gamma_c=\gamma_{c'}=10^{-4}\gamma_{ab}$.  We have
assumed the dephasings vanish, $\gamma_{jk}^{ph}=0$.}
\end{figure}

\begin{figure}
\includegraphics[width=1.0\columnwidth]{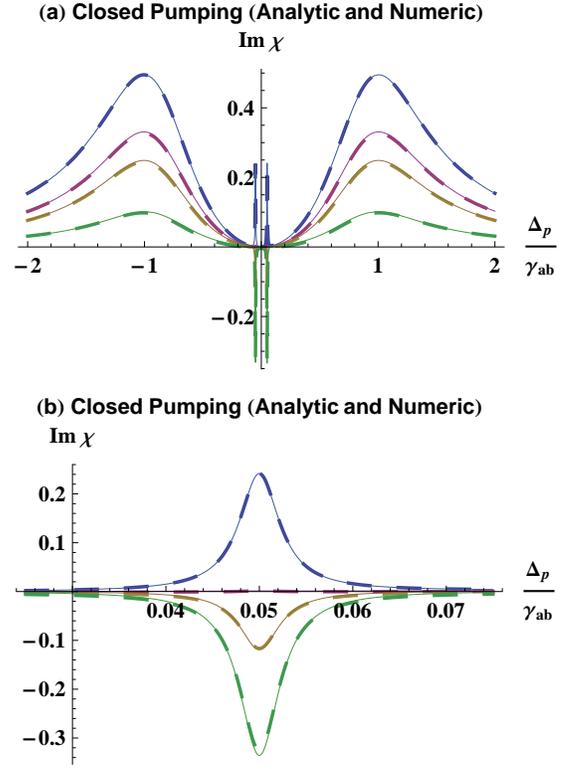}
\caption{\label{fig-closedImaginaryParts} (Color online) These are
analytic and numeric solutions for the imaginary part of the
susceptibility in the closed pumping case.  Again, the numeric solutions
are represented by wide dashed lines, and the analytics by narrower solid
lines.  The top plot shows the full spectrum as we vary $r$; the bottom
plot show a close up of one of the narrow features.  In both plots, moving
from the top curve to the bottom, the parameters are $r=0$ (blue lines),
$r=.005\gamma_{ab}$ (red lines), $r=.01\gamma_{ab}$ (brown lines), and
$r=.04\gamma_{ab}$ (green lines).    In all cases,
$\Omega_b=\Omega_c=.1\gamma_{ab}$, $\Omega_{\mu}=2\gamma_{ab}$, and
$\gamma_C=\gamma_{C'}=\gamma_{cc'}=\gamma_{b'b}=10^{-4}\gamma_{ab}$.}
\end{figure}

In Figs. \ref{fig-openImaginaryParts} and \ref{fig-closedImaginaryParts},
we plot the imaginary part of the linear susceptibility, Eq.
\ref{fullSusceptibility}, for various choices of pumping rates, in both
open and closed cases.  In both cases, we compare our analytic results
with a direct numerical solution of full density matrix equations of
motion. We see that the additional levels manifest themselves as two
tunable resonances located inside of the EIT transparency window.   In
general for arbitrary $\Delta_b$, the new resonances are symmetrically
located about $\Delta_p=0$ at the locations $\Delta_p=\pm
\Omega_b^{\text{eff}}/2=\pm \sqrt{\Delta_b^2+\Omega_b^2}/2$. For
$\Omega_{\mu},\gamma_{ab}\gg \Omega_b,\Omega_c,\gamma_C,\gamma_{C'}$,
their shape is approximately Lorentzian, given by (for
$\Delta_{\mu}=\Delta_b=0$):
\begin{align}
\Im[\tilde{\chi}^{(1)}]&\approx\frac{\gamma_{ab}^2\Omega_c^2}{2\Omega_{\mu}^2}(\text{Re}(\mathfrak{P}_B)-\tilde{\rho}_{c'c'}^{\text{st}})\notag\\
&\;\;\times\left(\frac{\Omega_c^2/\Omega_{\mu}^2+\gamma_{C'}/\gamma_{ab}}{(\Delta_p\mp\Omega_b/2)^2 +(\gamma_{ab}(\Omega_c^2/\Omega_{\mu}^2+\gamma_{C'}/\gamma_{ab}))^2}\right) \label{narrowlorentzian}
\end{align}
in the vicinity $\Delta_p\approx \pm \Omega_b/2$.  Eq.
\ref{narrowlorentzian} shows that we can expect  absorption for
$\text{Re}(\mathfrak{P}_B)>\tilde{\rho}_{c'c'}^{\text{st}}$ and gain for
$\tilde{\rho}_{c'c'}^{\text{st}}>\text{Re}(\mathfrak{P}_B)$.  Note that
these conditions are both necessary and sufficient for absorption and
gain, respectively, which implies first that no population need occupy the
excited state in order for amplification to occur (thus, we find
amplification without inversion) and second that coherent pumping is not
necessary for this amplification to occur.  Populations in the appropriate
ground states alone are necessary.


In the case of nonzero $\gamma_{C'}$, the widths of the features,
\begin{equation}
\Gamma_n=\gamma_{ab}(\Omega_c^2/\Omega_{\mu}^2+\gamma_{C'}/\gamma_{ab}), \label{linewidth}
\end{equation}
is the sum of the `power broadening' term $\gamma_{ab}\Omega_c^2/\Omega_{\mu}^2$ and the dephasing rate for $|c'\rangle$ while the height is given by
\begin{equation}\label{feature_height}
\Im[\tilde{\chi}^{(1)}(\pm\Omega_b/2)]=\frac{\Omega_c^2\gamma_{ab}(\text{Re}(\mathfrak{P}_B)-\tilde{\rho}^{\text{st}}_{c'c'})}{2(\gamma_{ab}\Omega_c^2+\Omega_{\mu}^2\gamma_{C'})}.
\end{equation}
The dependence on the population $\tilde{\rho}^{\text{st}}_{c'c'}$ is manifest in this expression: the height of the ultranarrow features varies linearly with the difference in populations between $\ket{b}$ (or in general a function of the population of $\ket{b}$) and $\ket{c'}$.  When the population in $\ket{c'}$ becomes large, the sign of the Lorentzian reverses, and the absorption line becomes a gain line.

The populations $\tilde{\rho}_{c'c'}$ and $\text{Re}(\mathfrak{P}_B)$ vary with the pumping rates for each of the two pumping configurations.  In the open pumping case, the relationships are linear in the pumping rates $r_b$ and $r_{c'}$; the other states do not attain appreciable population.  We find gain when
\begin{equation}\label{openGain}
\frac{r_{c'}}{r_b}>\frac{(2\gamma_{c'a}\Omega_c^2+\gamma_{C'}\Omega_{\mu}^2)}{(\gamma_b+\gamma_{b'})\Omega_{\mu}^2}.
\end{equation}
The linear dependence on the pumping rate in the open pumping
configuration permits additional control over the shapes of the lines. As
can be seen in Eq. \ref{feature_height}, the heights of the features are
proportional to $\Omega_c^2$.  In the case where $\Omega_c$ becomes small
(as is necessary to narrow the widths of the features), it is
theoretically possible to counteract the corresponding suppression of the
feature's height by increasing the pumping rates $r_b$ (for absorption) or
$r_{c'}$ (for gain).

\begin{figure}
\includegraphics[width=1.0\columnwidth]{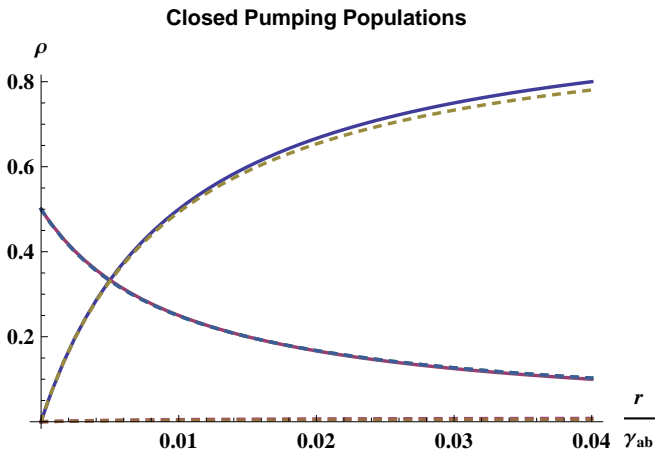}
\caption{\label{fig-populations} (Color online)  The analytic solutions (solid lines) for the populations in the closed pumping case, plotted with numerical solutions (dashed lines) to the full system of equations.   The lines with positive slope are the population of $\ket{c'}$ in each case; the other non-zero lines are the populations of $\ket{b},\ket{b'}$.  The dashed lines with vanishing population correspond to $\ket{a}$ and $\ket{c}$.   In all cases, $\Omega_b=\Omega_c=.1\gamma_{ab}$, $\Omega_{\mu}=2\gamma_{ab}$, and $\gamma_C=\gamma_{C'}=\gamma_{cc'}=\gamma_{b'b}=10^{-4}\gamma_{ab}$.}
\end{figure}

In the closed pumping configuration, meanwhile, the populations depend on a single parameter, $r$.  Now we find gain for
\begin{equation}\label{closedGain}
r>\frac{2\alpha_b\gamma_{c'a}\Omega_c^2}{\alpha_c\Omega_{\mu}^2}.
\end{equation}
Here, it is useful to compare the analytic expressions for the populations
to the populations found by direct numerical calculation (see Fig.
\ref{fig-populations}). As we see, there is excellent agreement in the
case of small $r$, as assumed. As $r$ grows, both plots plateau, but there
is a small deviation between the numerics and analytics.  This arises
because we assumed $r\ll\Omega_b,\Omega_c$.  Indeed, it is surprising that
agreement is acceptable for $r\gtrsim \Omega_b,\Omega_c$ in these plots.

\begin{figure}
\includegraphics[width=1.0\columnwidth]{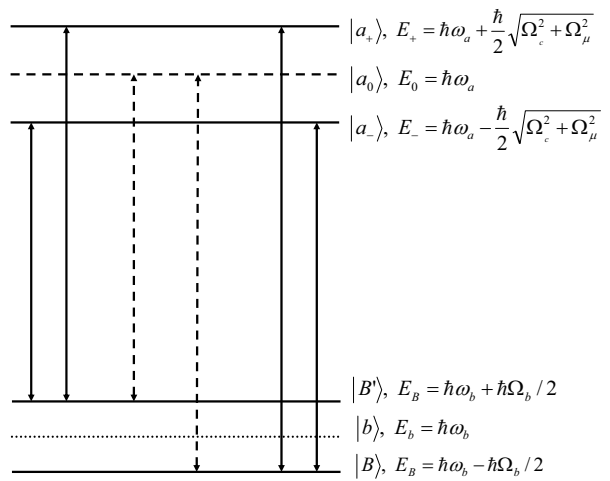}
\caption{\label{fig-dressedstate} Energy level diagram that indicates transitions induced by the probe laser between the ground state manifold $\{ |B\rangle, |B'\rangle \}$ and the excited state manifold $\{ |a_+\rangle, |a_-\rangle, |a_0\rangle \}$. Transitions to the gain state $|a_0\rangle$ are indicated by dashed lines. The energy of the bare state $|b\rangle$ is also shown for reference.}
\end{figure}

In the case where $\tilde{\rho}_{c'c'}$ becomes large, the presence of
gain lines without population inversion can be understood in terms of
dressed states.  Again assuming $\Delta_c=0$ and $\Delta_{\mu}=0$, the
eigenstates of the Hamiltonian for the
$\{|a\rangle,|c\rangle,|c'\rangle\}$ subsystem,
$H'=\hbar\omega_a(\ket{a}\bra{a}+\ket{c}\bra{c}+\ket{c'}\bra{c'})-(\hbar/2)(\Omega_{\mu}\ket{a}\bra{c}+\Omega_c
\ket{c'}\bra{c} +h.c)$, are
\begin{eqnarray}
|a_+\rangle &=&\frac{1}{\sqrt{2}}\left(\sin\theta|a\rangle+|c\rangle+\cos\theta|c'\rangle \right) \\
|a_-\rangle &=&\frac{1}{\sqrt{2}}\left(\sin\theta|a\rangle-|c\rangle+\cos\theta|c'\rangle \right) \\
|a_0\rangle &=& \cos\theta |a\rangle -\sin\theta|c'\rangle
\end{eqnarray}
where $\tan\theta=\Omega_{\mu}/\Omega_c$. The energies of the states
$|a_{\pm}\rangle$ are $E_{\pm}=\hbar\omega_a \pm
\hbar\sqrt{\Omega_{\mu}^2+\Omega_c^2}/2$ while $|a_0\rangle$ has energy
$E_0=\hbar\omega_a$. Note that $\{\ket{a},\ket{c},\ket{c'}\}$ is
isomorphic to a $\Lambda$ atom and so $\ket{a_0}$ is a dark state of the
sort familiar from STIRAP and coherent population trapping.  To
distinguish it from the more familiar dark state composed of $\ket{b}$ and
$\ket{c}$ responsible for EIT, we will call $\ket{a_0}$ the gain state.
Fig. \ref{fig-dressedstate} shows a schematic diagram of the energy levels
of the dressed ground state manifold $\{|B\rangle, |B'\rangle \}$, which
are coupled to all three states of the excited state manifold $\{
|a_+\rangle, |a_-\rangle, |a_0\rangle \}$ via the probe.

The eigenstate $|a_0\rangle$ is decoupled from the control laser, so there
will not be any destructive quantum interference in the probe absorption
or emission for transitions to $|a_0\rangle$.  This explains why we find
spectral lines at the locations corresponding to transitions from
$\ket{B}$ and $\ket{B'}$ to $\ket{a_0}$.  Moreover, in the limit as
$\Omega_c\ll\Omega_{\mu}$, we find $\sin\theta \gg \cos\theta$, and so
$\ket{a_0}\approx\ket{c'}$.  Thus in the dressed state basis, the
population in $\ket{c'}$ corresponds to large population in the gain
state.  The gain state energy remains $\hbar\omega_a$, however, even in
this limit, which means that although there is no population inversion in
the bare state basis, there is large inversion in this dressed state
basis.  Whereas when the population in $\ket{c'}$ is small, transitions
from the $\{ |B\rangle, |B'\rangle \}$ manifold to $|a_0\rangle$ lead to
absorption resonances at $\omega_a-\omega_{B,B'}$. These change to gain
lines as the population in $\ket{c'}\approx\ket{a_0}$ increases.

It is worth emphasizing that the gain lines described here arise from
different physical process than, say, standard driven two level pump-probe
spectroscopy. There, gain arises from the exchange of light quanta between
the strong pump field and the weak probe.  The dressed states used to
describe and explain pump-probe spectroscopy necessarily involve the
probed transition.  Here, the basis in which the gain state emerges does
not include the probed transition.  In the present system, the gain is due
to stimulated emission from a metastable quantum state, $\ket{a_0}$, to a
decoupled (to zeroth order in $\Omega_p/\Omega_{\mu}$) ground state.  It
is the coherent preparation of the $\{\ket{a},\ket{c},\ket{c'}\}$ manifold
that permits emission from this state (essentially $\ket{c'}$, in the
limits we have considered) even though the population of the bare excited
state is negligible.

The mechanism here also differs from textbook amplification without
inversion, as presented for instance by Scully and Zubairy \cite{Scully}.
There gain occurs because the coherence between the ground states of the
$\Lambda$ atom ($\ket{b}$ and $\ket{c}$, here) cancels absorption, leaving
only stimulated emission.  Thus standard inversionless amplification
system requires coherent pumping.  This permits gain even in the presence
of small population of the excited state, $\ket{a}$.  But crucially,
\emph{some} population in the excited state is necessary.  In the present
system, $\ket{a}$ may have zero population, so long as $\ket{c'}$ is
populated. We note that there are other approaches to amplification
without inversion (which is ultimately a term that describes systems
exhibiting a certain property, rather than the name of a particular class
of systems) that are closer in spirit to the current system
\cite{Mompart+Corbalan}.  But it is nonetheless worthwhile to distinguish
the DIGS approach from the ``standard'' amplification without inversion,
since their mutual connection to $\Lambda$ type atoms may produce
confusion.

\subsection{Re($\chi$): Anomalous dispersion}\label{dispersion}

\begin{figure}
\includegraphics[width=1.0\columnwidth]{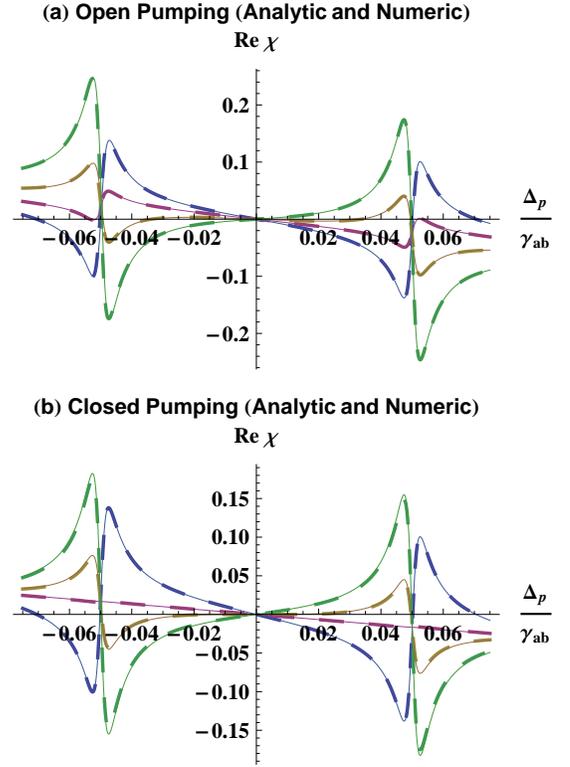}
\caption{\label{realPart}(Color online) Here we compare our analytic and
numeric solutions for various pumping rates in the open configuration (a)
and closed configuration (b) in the vicinity of the new features.  In both
cases, we see that increasing the pumping (or, increasing pumping to
$\ket{c'}$) results in a change from normal dispersion (negative slope, on
our sign conventions) to anomalous dispersion (positive slope) in the
region around zero detuning.  Once again, the analytic solutions are
represented by narrow solid lines, and the numerics by wide dashed lines.
The parameters in both cases are the same as in the close-ups in Figs.
\ref{fig-openImaginaryParts} (b) and \ref{fig-closedImaginaryParts} (b).
In particular, for (a) we have (from most negative slope to most
positive), $r_{c'}=0$ (blue lines), $r_{c'}=.002\gamma_{ab}$ (red lines),
$r_{c'}=.004\gamma_{ab}$ (brown lines), and $r_{c'}=.007\gamma_{ab}$
(green lines).  For (b), we have $r=0$ (blue lines), $r=.002\gamma_{ab}$
(red lines), $r=.004\gamma_{ab}$ (brown lines), and $r=.007\gamma_{ab}$
(green lines).}
\end{figure}

When the pumping rate is such that the gain peaks are large we find a
region of anomalous dispersion but low absorption.  This result should be
unsurprising, as the Kramers-Kronig relations guarantee that any pair of
sufficiently narrow and tall gain lines will give rise to anomalous
dispersion, given the analyticity of the susceptibility
\cite{Steinberg+Chiao}.  The dispersion is proportional to the first
derivative of the linear susceptibility.  In the present case, the
heights, widths, and separation of the peaks, and thus the magnitude and
range of the anomalous dispersion, are all tunable by varying $r$ (or
$r_b$ and $r_{c'}$), $\Omega_b$, and $\Omega_c$.  The real part of the
susceptibility is shown for several values of the pumping rate in Fig.
\ref{realPart}.  Our definition of the probe
detuning,\[\Delta_p=\omega_a-\omega_b-\nu_p,\] implies that the dispersion
is anomalous when the slope of the real part of $\chi$ is \emph{positive}.

\begin{figure}
\includegraphics[width=1.0\columnwidth]{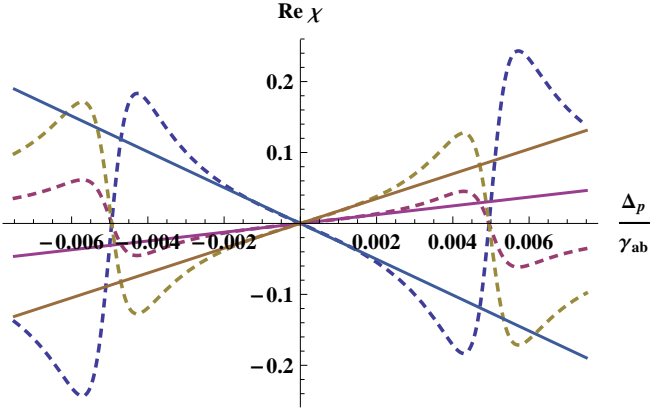}
\caption{\label{linearComp} (Color online) Here we compare our linear
approximation of the real part of the suseptibility, Eq.
\ref{dispersionEQ}, (solid lines) with the general analytic solution, Eq.
\ref{fullSusceptibility} (dashed lines) for various populations.  For
complete generality, and because the purpose is to show agreement between
the linear approximation and the full susceptibility, we consider the
populations directly, here, rather than limiting ourselves to one pumping
scheme or another.   In order from most negative to most positive slopes,
we have $\mathfrak{P}_B=1$ and $\tilde{\rho}_{c'c'}=0$ (blue lines);
$\mathfrak{P}_B=.25$ and $\tilde{\rho}_{c'c'}=.5$ (red lines); and
$\mathfrak{P}_B=.1$ and $\tilde{\rho}_{c'c'}=.8$ (brown lines).  In all
plots, $\gamma_C=\gamma_{C'}=\gamma_{ab}\times10^{-4}$,
$\Omega_{\mu}=2\gamma_{ab}$, $\Omega_b=\Omega_c=\gamma_{ab}/10$.}
\end{figure}

Near the bare transition frequency, $\Delta_p=0$, the real part of the susceptibility is approximately linear.  To second order in $\Omega_b/\Omega_{\mu}$ and $\Omega_c/\Omega_{\mu}$ (the order to which the solution for the susceptibility is valid), we find,
\begin{equation}\label{dispersionEQ}
\text{Re}(\tilde{\chi}^{(1)})=\frac{4\gamma_{ab}(\Omega_c^2\tilde{\rho}_{c'c'}-\text{Re}(\mathfrak{P}_B)(\Omega_b^2+\Omega_c^2))}{\Omega_b^2\Omega_{\mu}^2}\Delta_p.
\end{equation}
See Fig. \ref{linearComp} for a comparison of this linear solution and the full real part of the susceptibility for several populations.  This relation can be rewritten to reflect the relative populations necessary for dispersion to become anomalous.  The dispersion is anomalous just in case
\begin{equation}\label{dispersionRatio}
\frac{\tilde{\rho}_{c'c'}}{\text{Re}(\mathfrak{P}_B)}>1+\frac{\Omega_b^2}{\Omega_c^2}.
\end{equation}
Eq. \ref{dispersionRatio} can be rewritten as a constraint on $r_b$ and $r_{c'}$ in the opening pumping case and $r$ in the closed pumping case by substituting the expressions for the populations derived in sections \ref{open} and \ref{closed}.  In the open pumping case, the constraint on $r_{c'}/r_b$ is given by
\begin{equation}\label{openDispersion}
\frac{r_{c'}}{r_b}>\frac{(2\gamma_{c'a}\Omega_c^2+\gamma_{C'}\Omega_{\mu}^2)(\Omega_b^2+\Omega_c^2)}{(\gamma_b+\gamma_{b'})\Omega_{\mu}^2\Omega_{c}^2},
\end{equation}
whereas the corresponding constraint on $r$ in the closed pumping case is given by,
\begin{equation}\label{closedDispersion}
r>\frac{2\alpha_b\gamma_{c'a}(\Omega_b^2+\Omega_c^2)}{\alpha_c\Omega_{\mu}^2}.
\end{equation}

It is tempting to consider the limit that
$\Omega_b^2/\Omega_c^2\rightarrow0$.  However, Eqs. \ref{dispersionEQ} and
\ref{dispersionRatio} are not valid in this limit. For one, taking
$\Omega_b\rightarrow0$ makes sense in the context of Eq.
\ref{dispersionRatio}, but not Eq. \ref{dispersionEQ}.  More importantly,
the assumption of linearity only holds when there is a transparency window
between the two peaks, or roughly when
$\Omega_c^2/\Omega_{\mu}^2\ll\Omega_b/\gamma_{ab}$, which corresponds to
the case where the separation of the peaks is larger than their widths.
It follows that, while inversion of the populations of
$\ket{a_0}\approx\ket{c'}$ and $\ket{b}$ is sufficient to produce gain
lines, the constraint on the population of $\ket{a_0}$  necessary for
anomalous dispersion is more stringent.  This result can be seen clearly
in comparing Eqs. \ref{openDispersion} and \ref{closedDispersion} to the
corresponding expressions in section \ref{gain}, Eqs. \ref{openGain} and
\ref{closedGain}.  The pumping rate necessary for producing anomalous
dispersion increases as the square of the width of the window between the
peaks, supposing that the widths of the peaks are held constant.  Thus,
though it is in principle possible to produce wide spectral regions of
anomalous dispersion, there is a practical barrier imposed by how rapidly
one can pump atoms into $\ket{c'}$.

\begin{figure}
\includegraphics[width=1.0\columnwidth]{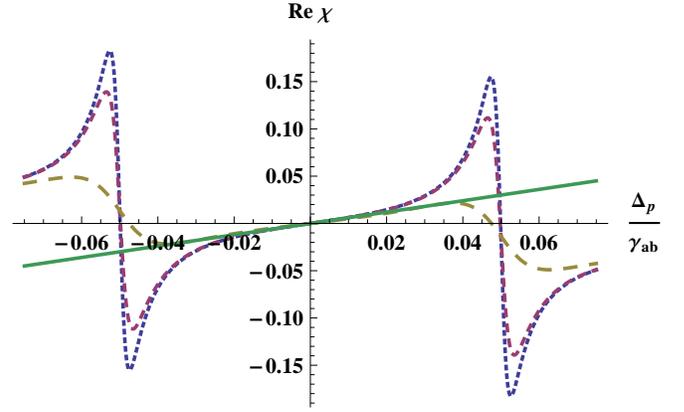}
\caption{\label{dephasing} (Color online) Here we compare our linear
approximation of the real part of the suseptibility, Eq.
\ref{dispersionEQ}, with the general analytic solution, Eq.
\ref{fullSusceptibility}, as dephasing increases.  The (blue) dotted line
has $\gamma_C=\gamma_{C'}=\gamma_{ab}\times 10^{-4}$; the (red) dashed
line has $\gamma_C=\gamma_{C'}=\gamma_{ab}\times 10^{-3}$; the (yellow)
broad-dashed line has $\gamma_C=\gamma_{C'}=\gamma_{ab}/100$.  The solid
(green) line is the linear approximation.  Note that although the
susceptibility is depressed in the vicinity of the new resonances as the
dephasing increases, the dispersion in the linear regime does not change.
In all plots, we take the generalized populations $\tilde{\rho}_{c'c'}=.8$
and $\text{Re}(\mathfrak{P}_B)=.1$, while other parameters are
$\Omega_{\mu}=2\gamma_{ab}$, $\Omega_b=\Omega_c=\gamma_{ab}/10$.}
\end{figure}

A distinctive and important feature of Eq. \ref{dispersionEQ} is that the
dispersion does not depend on the dephasings $\gamma_C$ and $\gamma_{C'}$.
This result is not a relic of the approximations that went into deriving
the linearized equation, as can be seen in Fig. \ref{dephasing}.  The
effect of dephasing between $\ket{c'}$ and $\{\ket{b},\ket{b'}\}$ is to
destroy the coherences responsible for producing the new narrow
resonances. But even as the resonances shrink, the region between the
peaks near zero detuning remains unchanged, until $\gamma_{C'}\gg\Omega_c$
and the resonances vanish altogether.

The values of the anomalous dispersion that we predict have dramatic
consequences for group velocity, which in the current context can be
written $v_g=c/n_g$, where $n_g$ is the group velocity
index,
\[n_g=n-\frac{\nu_p}{2n}\frac{\partial\Re(\chi^{(1)})}{\partial\Delta_p}.\]
$n=(1+\Re(\chi^{(1)}))^{1/2}$ is the index of refraction, as defined
above.  The absolute value of the dispersion in the anomalous regime, for
large population $\ket{c'}$, is comparable to the magnitude of EIT
dispersion, as can be seen by examining Eq. \ref{dispersionEQ}.  In the
case where $\mathfrak{P}_B=1/2$, the dispersion becomes approximately that
of ``standard EIT,'' which we take to be EIT in an identical system,
without the two additional fields/levels we have introduced here.
\begin{equation}\label{dispersionNoPumping}
\frac{\partial\Re(\tilde{\chi}^{(1)})}{\partial\Delta_p}=- \frac{4\gamma_{ab}\text{Re}(\mathfrak{P}_B)(\Omega_b^2+\Omega_c^2)}{\Omega_b^2\Omega_{\mu}^2}.
\end{equation}
In the opposite limit, of very large pumping, $\tilde{\rho}_{c'c'}\rightarrow 1$ and we find
\begin{equation}\label{dispersionAllPumping}
\frac{\partial\Re(\tilde{\chi}^{(1)})}{\partial\Delta_p}= \frac{4\gamma_{ab}\Omega_c^2\tilde{\rho}_{c'c'}}{\Omega_b^2\Omega_{\mu}^2}
\end{equation}
For $\Omega_c=\Omega_b$, Eq. \ref{dispersionAllPumping} reduces to the
expression for EIT dispersion, with opposite sign. This means that if one
begins with a suitable EIT system and then introduces the additional
couplings and pump processes, one can generate negative group velocities
of the same magnitude as the ultraslow light observed by \cite{Hau+etal},
\cite{Kash+etal}, and \cite{Budker+etal}.

\begin{figure}
\includegraphics[width=1.0\columnwidth]{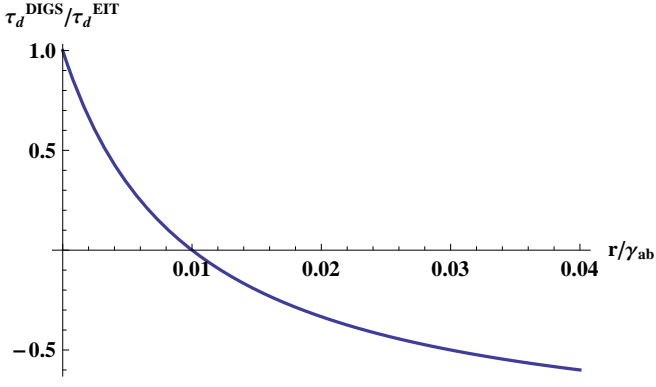}
\caption{\label{groupDelay}   The ratio
$\tau_d^{\text{DIGS}}/\tau_d^{\text{EIT}}$ as a function of the closed
pumping rate, $r/\gamma_{ab}$.  We see that for $r>\gamma_{ab}/10$, the
time delay becomes negative; for $r>3\gamma_{ab}/10$, the magnitude of the
negative time delay comes within a factor of 2 of the EIT time delay,
permitting much faster light than previously observed. Here
$\Omega_b=\Omega_c=0.1\gamma_{ab}$}
\end{figure}

Negative group velocities can best be understood in terms of the group
delay,
\[\tau_d=\ell(1/v_g-1/c)\]
where $\ell$ is the sample thickness. To see the point made at the end of
the last paragraph most clearly, suppose that a given EIT system exhibits
a group delay of $\tau_d^{\text{EIT}}$.  Then, for index of refraction
$n\approx 1$ (as we have here) we can expect a group delay in the same
system prepared with the additional DIGS couplings of,
\begin{equation}
\tau_d^{\text{DIGS}}=\tau_d^{\text{EIT}}\frac{\Re(\mathfrak{P}_B)(\Omega_b^2+\Omega_c^2)-\Omega_c^2\tilde{\rho}_{c'c'}}{\Omega_b^2}.
\end{equation}
We plot $\tau_d^{\text{DIGS}}/\tau_d^{\text{DIGS}}$ as a function of the closed pumping rate in Fig. \ref{groupDelay}.

To be perfectly concrete, take as a sample system the $^{87}$Rb vapor cell
prepared in \cite{Kash+etal}, modified to include the additional couplings
(see section \ref{conclusion}).  We can assume that
$\Omega_b\approx\Omega_c$, so that the details of the strengths of the
couplings are irrelevant.  They observe group velocities of $90$m/s
through their 2.5cm long sample, corresponding to a group delay of .26ms.
In the corresponding DIGS system, with pumping such that
$\tilde{\rho}_{c'c'}\approx.8$ and $\mathfrak{P}_B\approx .1$ (this
corresponds to a closed pumping rate of approximately $r=.04\gamma_{ab}$
for $\Omega_b\approx\Omega_c\approx .1\gamma_{ab}$ and
$\Omega_{\mu}=2\gamma_{ab}$), we should expect a group velocity of
-150m/s, corresponding to a group delay of -.156ms and a group velocity
index of $n_g=-2\times 10^6$.  These numbers represent an improvement of
four orders of magnitude over Ref. \cite{WKD-Nature}, who found a group
velocity index of $n_g\approx-310$, and of two orders of magnitude over
Ref. \cite{Kim+etal} who observed a group velocity index of $n_g=-14\;
400$ in a Cs atomic vapor system.  This latter result is the largest
superluminal group velocity yet observed directly.

\section{Doppler Broadening}\label{doppler}

Doppler broadening is an important experimental constraint on the current
system. Thus far, we have disregarded Doppler broadening in our treatments
of DIGS systems, and so we will focus on it here.  For a single photon
with wave vector $\vec{k}$ incident on an atom moving with velocity
$\vec{v}$, the Doppler effect shifts the atomic transition frequency
$\omega_0$ by
\begin{equation*}
\omega_D=\omega_0+\vec{k}\cdot\vec{v}.
\end{equation*}
The ``Doppler width,'' then, is given by the width of the atomic velocity
distribution
$\sigma_D=\langle(\omega_D-\omega_0)^2\rangle=\omega_0\sigma_v/c$. For a
Maxwell-Boltzmann distribution $\sigma_v$ is given by (FWHM)
\begin{equation*}
\sigma_v=2\sqrt{2\ln2\frac{k_BT}{m}}
\end{equation*}
where $k_B$ is the Boltzmann constant, $T$ is the temperature, and $m$ is the mass of the atoms.

From these considerations, we conclude that Doppler shifts will be
unimportant for the RF fields. Working, for instance, at room temperature
$(T=300$K$)$, in a gas of Rubidium atoms, we would find that
$\sigma_v\approx 400$m/s.  For an RF field, we can take $\omega_0\sim
100$MHz, which gives an estimation of the Doppler width at $\sigma_D\sim
100$Hz. Thus for the RF fields, the effect of Doppler broadening will be
much less than the homogeneous broadening from dephasing (which we have
thus far estimated in our plots at
$\gamma_b,\gamma_{b'},\gamma_C,\gamma_{C'}\approx10^{-4}\gamma_{ab}\approx10^3$Hz).
In the same gas, however, taking $\omega_0\approx 1$PHz---a characteristic
frequency of light---we find that the Doppler width will be approximately
$\sigma_D\approx 1$GHz$\gg \gamma_{ab}$. Thus Doppler broadening will
dominate the optical transitions
$\{\ket{b},\ket{b'}\}\leftrightarrow\ket{a}$ and
$\{\ket{c},\ket{c'}\}\leftrightarrow\ket{a}$.  In a laser cooled system,
where $T\approx1\mu $K$-100\mu $K, the Doppler width would be reduced to
approximately $\gamma_{ab}/10-\gamma_{ab}/100$.

We can model the Doppler broadening by averaging over a Gaussian distribution of the one photon probe detuning $\Delta_p$ and the two photon detuning, $\delta=\Delta_p-\Delta_{\mu}$.  Written in terms of these two detunings (see appendix D for a full statement of the susceptibility including all nonzero detunings) 
the Doppler broadened susceptibility as a function of the mean probe
detuning, $\Delta_p$, with mean pump detuning fixed at 0 is given by
\begin{align}\label{dopplerBroadening}
\tilde{\chi}^{(1)}_D(\Delta_p)&=\frac{1}{2\pi\sigma_{\Delta_p}\sigma_{\delta}}\int_{-\infty}^{\infty}\int_{-\infty}^{\infty}d\Delta_p'd\delta \;\; \tilde{\chi}^{(1)}(\Delta_p',\delta)\notag\\
&\;\;\times e^{-(\Delta_p'-\Delta_p)^2/(2\sigma_{\Delta_p}^2)}e^{-(\delta-\Delta_p)^2/(2\sigma_{\delta}^2)}.
\end{align}

\begin{figure}
\includegraphics[width=1.0\columnwidth]{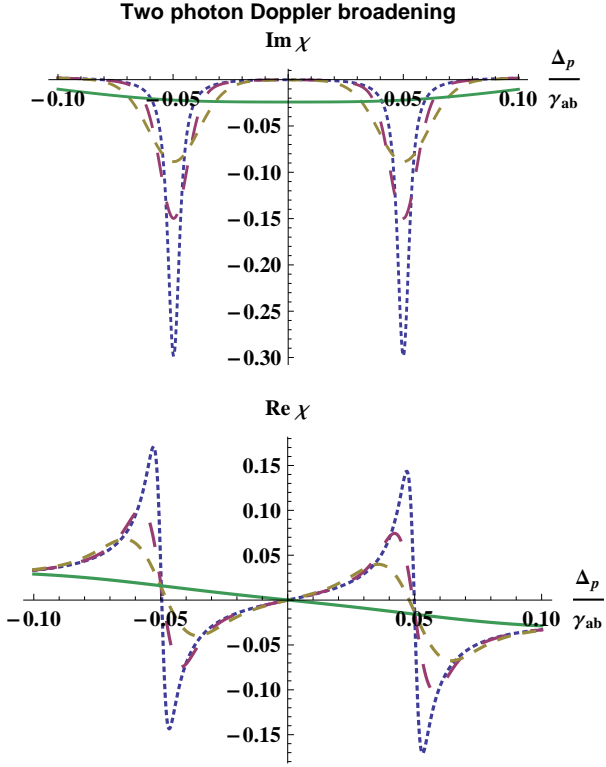}
\caption{\label{twoPhoton} (Color online) Here we have the two photon
Doppler effect on the real and imaginary parts of the susceptibility near
the narrow features.  In both plots, we keep the one photon variance
constant (to clarify the effect of the two photon broadening) at
$\sigma_{\Delta_p}=.001\gamma_{ab}$.  The formatting of the lines is
consistent between the two plots.  In both, the blue dotted line
corresponds to $\sigma_{\delta}=.001\gamma_{ab}$, the red dashed line has
$\sigma_{\delta}=.005\gamma_{ab}$, the brown dashed line has
$\sigma_{\delta}=.01\gamma_{ab}$, and the solid green line (for which the
features have vanished altogether) has $\sigma_{\delta}=.05\gamma_{ab}$.
For simplicity and generality, we have used generalized populations rather
than particular models for open or closed pumping, with
$\mathfrak{P}_B=.1$ and $\tilde{\rho}_{c'c'}=.8$, and we have allowed the
dephasings $\gamma_C$ and $\gamma_{C'}$ to vanish.  The other parameters
are $\Omega_{\mu}=2\gamma_{ab}$, $\Omega_b=\Omega_c=\gamma_{ab}/10$.}
\end{figure}

\begin{figure}
\includegraphics[width=1.0\columnwidth]{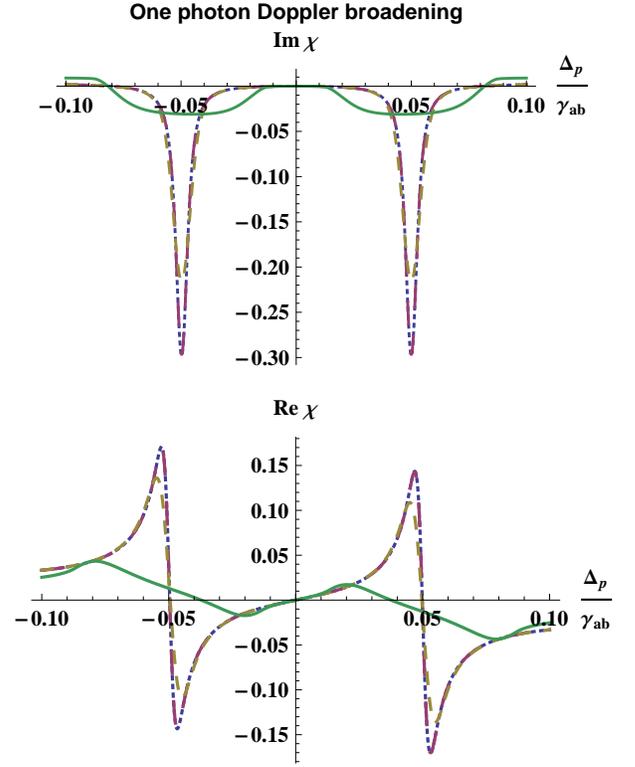}
\caption{\label{onePhoton}(Color online)  The one photon Doppler effect.
Again, we keep the two photon broadening constant at
$\sigma_{\delta}=.001\gamma_{ab}$.  In this case we show a wide array of
values for $\sigma_{\Delta_p}$, as the features are much less sensitive to
the one photon broadening than the two photon broadening---indeed, each
line represents variance an order of magnitude larger than the last. In
both the top and bottom plots, the blue dotted line corresponds to
$\sigma_{\Delta_p}=.01\gamma_{ab}$, the red dashed line has
$\sigma_{\Delta_p}=.1\gamma_{ab}$, the brown dashed line has
$\sigma_{\Delta_p}=\gamma_{ab}$, and the solid green line has
$\sigma_{\Delta_p}=10\gamma_{ab}$.  For simplicity and generality, we have
once again used generalized populations rather than particular models for
open or closed pumping, with $\mathfrak{P}_B=.1$ and
$\tilde{\rho}_{c'c'}=.8$, and we have allowed the dephasings $\gamma_C$
and $\gamma_{C'}$ to vanish.  The other parameters are
$\Omega_{\mu}=2\gamma_{ab}$, $\Omega_b=\Omega_c=\gamma_{ab}/10$.}
\end{figure}

Evaluating Eq. \ref{dopplerBroadening} numerically for a variety of
variances (see Figs. \ref{twoPhoton} and \ref{onePhoton}), we can draw
several conclusions about the constraints imposed by Doppler broadening.
For one, the narrow features are unaffected by the single photon
broadening, at least while $\sigma_{\Delta_p}/\gamma_{ab}\lesssim 1$.  To
see why this would be, consider the expansion of Im$(\tilde{\chi}^{(1)})$
around $\pm\Omega_b/2$ in terms of the one and two photon detunings.  We
find,
\begin{align}\label{moreGeneralLorentzian}
&\Im(\tilde{\chi}^{(1)}(\Delta_p,\delta))\approx\frac{\gamma_{ab}\Omega_c^2}{2\Omega_{\mu}^2}(\text{Re}(\mathfrak{P}_B)-\tilde{\rho}_{c'c'}^{\text{st}})\times\notag\\
&\left(\frac{\gamma_{ab}\Omega_c^2/\Omega_{\mu}^2+\gamma_{C'}}{(\gamma_{ab}\Omega_c^2/\Omega_{\mu}^2+\gamma_{C'})^2+(\delta\mp\Omega_b/2+\Omega_c^2/\Omega_{\mu}^2(\Delta_p\mp\Omega_b/2))^2}\right)
\end{align}
Under the assumptions that $\Omega_c^2/\Omega_{\mu}^2\ll1$ and
$\Delta_{\mu}\rightarrow0$, this expression reduces to the Lorentzian
given in Eq. \ref{narrowlorentzian}. In this case, Eq.
\ref{moreGeneralLorentzian} depends on $\delta$ alone (because the term
$\Omega_c^2/\Omega_{\mu}^2(\Delta_p\mp \Omega_b/2)$ can be neglected in
the denominator), which explains why the widths of the features are more
sensitive to inhomogeneous broadening of the two photon transition.
Meanwhile, when $\sigma_{\Delta_p}\Omega_c^2/\Omega_{\mu}^2$ approaches
the width of the features, broadening of the single photon process becomes
important.  Nevertheless, even for
$\sigma_{\Delta_p}/\gamma_{ab}\approx10$, our numerics show that the
features will persist, albeit in suppressed form.

The narrow features are far more sensitive to the two photon broadening.
When $\sigma_{\delta}$ approaches the widths of the features,
$\gamma_{ab}\Omega_c^2/\Omega_{\mu}^2+\gamma_{C'}$, they are rapidly
broadened and ultimately washed out.  One possible way of avoiding this
difficulty is to work with a Doppler-free geometry, where the probe and
the control lasers are copropagating.  In this case, the two photon
broadening vanishes and it would be possible to observe the (one photon
broadened) narrow resonances even in a room temperature system, as per the
discussion in the previous paragraph.  It would also be possible to work
in a system with two photon Doppler broadening, so long as the temperature
is sufficiently low.  For $\Omega_c^2/\Omega_{\mu}^2\approx.01$, it would
be possible to work in a laser cooled system; for smaller
$\Omega_c/\Omega_{\mu}$, a quantum degenerate system would be necessary to
realize the gain.

The anomalous dispersion, meanwhile, is more robust under Doppler
broadening.  Just as when one increases the homogeneous broadening of the
narrow lines, the slope of the real part of the susceptibility in the
anomalous regime is unchanged even for large inhomogeneous broadening.  It
is only when the features are washed out entirely by the inhomogeneous
broadening that the anomalous dispersion is lost.  In particular, the
anomalous dispersion is not reduced when the variance in the one photon
detuning, $\sigma_{\Delta_p}$, is as large as $10\gamma_{ab}$.  The
dispersion is more sensitive to the two photon detuning, but only because
the features themselves vanish entirely for a smaller value of
$\sigma_{\delta}$.  This means that the anomalous dispersive regime should
be readily observable and unsuppressed even in a room temperature gas,
provided one uses a Doppler-free geometry to eliminate the two photon
broadening.  Our numerical calculation predicts that even in the absence
of co-propagating lasers, the dispersion will be unchanged for
$\sigma_{\delta}\lesssim\gamma_{ab}/100$, even though the resonances will
be severely broadened.  Thus it should be possible to observe the
anomalous dispersion in a laser-cooled gas without a Doppler-free
geometry.

\section{Summary and Conclusions}\label{conclusion}

As a final note on experimental realizability, we propose several possible
level structures. The levels we propose here are for $^{87}$Rb, though the
equivalent levels in $^{23}$Na would work equally well.  The first
proposal is a modification of the structure used in the EIT experiment in
Ref. \cite{Kash+etal}, which we used above to make numerical predictions
for the values of the negative group delay in our system.  For the closed
pumping system, one might use $D_1$ line with
$\ket{a}=\ket{5P_{1/2},F=2,m_f=2}$, $\ket{b}=\ket{5S_{1/2},F=1,m_f=1}$,
$\ket{b'}=\ket{5S_{1/2},F=1,m_f=0}$, $\ket{c}=\ket{5S_{1/2},F=2,m_f=1}$,
and $\ket{c'}=\ket{5S_{1/2},F=2,m_f=2}$. Alternatively, one might use the
$D_2$ line with $\ket{a}=\ket{5P_{3/2},F=2,m_f=2}$ and the same ground
states from the $5S_{1/2}$ manifold as just given for the $D_1$ line. Then
$\ket{a}$ decays only to $\ket{b}$, $\ket{c}$, and $\ket{c'}$, as
described by our model. (Note that choosing levels where $\ket{a}$ decays
also to $\ket{b'}$ will not change qualitatively any of our results
provided $\Omega_b \neq 0$ and $\Delta_b=0$.) For both of the
configurations we propose, both the probe and the control beam would have
to be $\sigma_+$ polarized.

In conclusion, we have studied the effects of various pumping
configurations on the linear response of a driven five-level atom. We have
found that when the population of one of the ground states, $\ket{c'}$,
becomes large, it produces two amplification resonances without population
inversion in the bare state basis.  We analyzed the dependence of the
population of $\ket{c'}$ on two different pumping configurations.
Moreover, we have shown that in the region between the two gain lines, it
is possible to tune the system to permit anomalous dispersion.  We have
studied the effects of Doppler broadening on this system, and concluded
that the anomalous dispersion is robust under both homogeneous and
inhomogeneous broadening of the gain lines, so long as the gain lines do
not vanish.

C.P.S. acknowledges support for this work from the National Science
Foundation award no. 0757933.

\appendix

\section{}
The equations of motion for the system described in section \ref{model} are given by,
\begin{subequations}\label{baseEquations}
\begin{align}
i\dtime{\tilde{\rho}_{aa}}&=
 -\frac{\Omega_{\mu}}{2}( \tilde{\rho}_{ca}-\tilde{\rho}_{ac})+\frac{\Omega_{p}}{2}(\tilde{\rho}_{ab}- \tilde{\rho}_{ba})\\
i\dtime{\tilde{\rho}_{bb}}&= \frac{\Omega_{b}}{2}(\tilde{\rho}_{bb'}-\tilde{\rho}_{b'b})-\frac{\Omega_{p}}{2}(\tilde{\rho}_{ab} - \tilde{\rho}_{ba})\\
i\dtime{\tilde{\rho}_{b'b'}}&= -\frac{\Omega_{b}}{2}(\tilde{\rho}_{bb'}-\tilde{\rho}_{b'b})\\
i\dtime{\tilde{\rho}_{cc}}&=  -\frac{\Omega_{c}}{2}(\tilde{\rho}_{c'c}-\tilde{\rho}_{cc'}) +\frac{\Omega_{\mu}}{2} (\tilde{\rho}_{ca}-\tilde{\rho}_{ac} ) \\
i\dtime{\tilde{\rho}_{c'c'}}&=  \frac{\Omega_{c}}{2}(\tilde{\rho}_{c'c}-\tilde{\rho}_{cc'})
\end{align}
and then the off-diagonals,
\begin{align}
i\dtime{\tilde{\rho}_{ab}}&= \Delta_p\tilde{\rho}_{ab} -\frac{\Omega_{\mu} }{2}\tilde{\rho}_{cb} +\frac{\Omega_{b}}{2}\tilde{\rho}_{ab'} +\frac{\Omega_{p}}{2}(\tilde{\rho}_{aa}- \tilde{\rho}_{bb}) \\
i\dtime{\tilde{\rho}_{ab'}}&= (\Delta_p-\Delta_b)\tilde{\rho}_{ab'} -\frac{\Omega_{\mu}}{2} \tilde{\rho}_{cb'} -\frac{\Omega_{p}}{2}\tilde{\rho}_{bb'} +\frac{\Omega_{b}}{2}\tilde{\rho}_{ab}\\
i\dtime{\tilde{\rho}_{ca}}&= -\Delta_{\mu}\tilde{\rho}_{ca} +\frac{\Omega_{p}}{2}\tilde{\rho}_{cb} -\frac{\Omega_{\mu}}{2}( \tilde{\rho}_{aa} -\tilde{\rho}_{cc} ) -\frac{\Omega_{c}}{2}\tilde{\rho}_{c'a}\\
i\dtime{\tilde{\rho}_{c'a}}&= (\Delta_c-\Delta_{\mu})\tilde{\rho}_{c'a} +\frac{\Omega_{\mu}}{2}\tilde{\rho}_{c'c} +\frac{\Omega_{p}}{2}\tilde{\rho}_{c'b}- \frac{\Omega_{c}}{2}\tilde{\rho}_{ca}\\
i\dtime{\tilde{\rho}_{cb}}&= (\Delta_p-\Delta_{\mu})\tilde{\rho}_{cb} +\frac{\Omega_{b}}{2}\tilde{\rho}_{cb'} +\frac{\Omega_{p}}{2}\tilde{\rho}_{ca}  -\frac{\Omega_{\mu}}{2}\tilde{\rho}_{ab}\notag\\
&\;\; -\frac{\Omega_{c}}{2}\tilde{\rho}_{c'b}
\end{align}
\begin{align}
i\dtime{\tilde{\rho}_{cb'}}&=(\Delta_p-\Delta_b-\Delta_{\mu})\tilde{\rho}_{cb'} +\frac{\Omega_{b}}{2}\tilde{\rho}_{cb} -\frac{\Omega_{\mu}}{2}\tilde{\rho}_{ab'} \notag\\
&\;\;-\frac{\Omega_{c}}{2}\tilde{\rho}_{c'b'}\\
i\dtime{\tilde{\rho}_{c'b}}&=(\Delta_p+\Delta_c-\Delta_{\mu})\tilde{\rho}_{c'b} +\frac{\Omega_{b}}{2}\tilde{\rho}_{c'b'} +\frac{\Omega_{p}}{2}\tilde{\rho}_{c'a}\notag\\
&\;\;- \frac{\Omega_{c}}{2}\tilde{\rho}_{cb}\\
i\dtime{\tilde{\rho}_{c'b'}}&= (\Delta_p-\Delta_b+\Delta_c-\Delta_{\mu})\tilde{\rho}_{c'b'} +\frac{\Omega_{b}}{2}\tilde{\rho}_{c'b} \notag\\
&\;\;- \frac{\Omega_{c}}{2}\tilde{\rho}_{cb'}\\
i\dtime{\tilde{\rho}_{bb'}}&= -\Delta_b\tilde{\rho}_{bb'} -\frac{\Omega_{p}}{2}\tilde{\rho}_{ab'}+\frac{\Omega_{b}}{2}(\tilde{\rho}_{bb}-\tilde{\rho}_{b'b'} )\\
i\dtime{\tilde{\rho}_{c'c}}&= \Delta_c\tilde{\rho}_{c'c} +\frac{\Omega_{\mu}}{2}\tilde{\rho}_{c'a} - \frac{\Omega_{c}}{2}(\tilde{\rho}_{cc}-\tilde{\rho}_{c'c'} )
\end{align}
\end{subequations}
We have defined detunings $\Delta_p=\omega_a-\omega_b-\nu_p$, $\Delta_{\mu}=\omega_a-\omega_c-\nu_{\mu}$, $\Delta_b=\omega_{b'}-\omega_b-\nu_b$, and $\Delta_c=\omega_{c'}-\omega_c-\nu_c$.

\section{}
We transform the equations of motion to first order in $\Omega_p/\Omega_{\mu}$ by diagonalizing the $\{\ket{b},\ket{b'}\}$ and $\{\ket{c},\ket{c'}\}$ subspaces of the Hamiltonian via the matrix
\begin{equation}
D=\begin{pmatrix}
1 &0 &0 &0 &0\\
0 &\cos\theta_b &\sin\theta_b &0 &0\\
0 &-\sin\theta_b &\cos\theta_b &0 &0\\
0 &0 &0 &\cos\theta_c &\sin\theta_c\\
0 &0 &0 &-\sin\theta_c &\cos\theta_c
\end{pmatrix}
\end{equation}
where
\begin{align*}
&\cos\theta_i=\sqrt{\frac{1+\Delta_i/\Omega_i^{\text{eff}}}{2}}& &\sin\theta_i=\sqrt{\frac{1-\Delta_i/\Omega_i^{\text{eff}}}{2}}&
\end{align*}
and
\begin{equation*}
\Omega_i^{\text{eff}}=\sqrt{\Delta^2_i+\Omega_i^2}.
\end{equation*}

In this basis, the Hamiltonian becomes
\begin{widetext}
\begin{equation}
D\tilde{\mathcal{H}}D^{\dagger}=\frac{\hbar}{2}
\begin{pmatrix}
2\omega_a &-\Omega_{p}\cos\theta_b &\Omega_p\sin\theta_b &-\Omega_{\mu}\cos\theta_c &\Omega_{\mu}\sin\theta_c\\
-\Omega_{p}\cos\theta_b &2\omega_b+\Delta_{b}+2\nu_p-\Omega_{b}^{\text{eff}} &0 &0 &0\\
\Omega_p\sin\theta_b &0 &2\omega_b+\Delta_{b}+2\nu_p+\Omega_{b}^{\text{eff}} &0 &0\\
-\Omega_{\mu}\cos\theta_c &0 &0 &2\omega_c+\Delta_{c}+2\nu_{\mu}-\Omega_{c}^{\text{eff}} &0\\
\Omega_{\mu}\sin\theta_c &0 &0 &0 &2\omega_c+\Delta_{c}+2\nu_{\mu}+\Omega_{c}^{\text{eff}}
\end{pmatrix}.
\end{equation}

Written, for now, without decay (which will require some approximations to transfer into this basis), we find
\begin{subequations}
\begin{align}
i\dtime{\tilde{\rho}_{aB}}&=(\Delta_p-\frac{\Delta_b}{2}+\frac{\Omega_b^{\text{eff}}}{2})\tilde{\rho}_{aB}- \frac{\Omega_p}{2}(\cos\theta_b\tilde{\rho}_{BB}-\sin\theta_b\tilde{\rho}_{B'B}) -\frac{\Omega_{\mu}}{2}(\cos\theta_c\tilde{\rho}_{CB}-\sin\theta_c\tilde{\rho}_{C'B}) +\frac{\Omega_p}{2}\cos\theta_b\tilde{\rho}_{aa}\\
i\dtime{\tilde{\rho}_{CB}}&=(-\Delta_{\mu}+\Delta_p-\frac{\Delta_{b}}{2}+\frac{\Omega_{b}^{\text{eff}}}{2} +\frac{\Delta_{c}}{2}-\frac{\Omega_{c}^{\text{eff}}}{2})\tilde{\rho}_{CB} +\frac{\Omega_p}{2}\cos\theta_b\tilde{\rho}_{Ca}-\frac{\Omega_{\mu}}{2}\cos\theta_c\tilde{\rho}_{aB}\\
i\dtime{\tilde{\rho}_{C'B}}&=(-\Delta_{\mu}+\Delta_p-\frac{\Delta_{b}}{2}+\frac{\Omega_{b}^{\text{eff}}}{2} +\frac{\Delta_{c}}{2}+\frac{\Omega_{c}^{\text{eff}}}{2})\tilde{\rho}_{C'B} +\frac{\Omega_p}{2}\cos\theta_b\tilde{\rho}_{C'a}+\frac{\Omega_{\mu}}{2}\sin\theta_c\tilde{\rho}_{aB}
\end{align}
\end{subequations}
and
\begin{subequations}
\begin{align}
i\dtime{\tilde{\rho}_{aB'}}&=(\Delta_p-\frac{\Delta_b}{2}-\frac{\Omega_b^{\text{eff}}}{2})\tilde{\rho}_{aB}- \frac{\Omega_p}{2}(\cos\theta_b\tilde{\rho}_{BB'}-\sin\theta_b\tilde{\rho}_{B'B'}) -\frac{\Omega_{\mu}}{2}(\cos\theta_c\tilde{\rho}_{CB'}-\sin\theta_c\tilde{\rho}_{C'B'}) -\frac{\Omega_p}{2}\sin\theta_b\tilde{\rho}_{aa}\\
i\dtime{\tilde{\rho}_{CB'}}&=(-\Delta_{\mu}+\Delta_p-\frac{\Delta_{b}}{2}-\frac{\Omega_{b}^{\text{eff}}}{2} +\frac{\Delta_{c}}{2}-\frac{\Omega_{c}^{\text{eff}}}{2})\tilde{\rho}_{CB'} -\frac{\Omega_p}{2}\sin\theta_b\tilde{\rho}_{Ca}-\frac{\Omega_{\mu}}{2}\cos\theta_c\tilde{\rho}_{aB'}\\
i\dtime{\tilde{\rho}_{C'B'}}&=(-\Delta_{\mu}+\Delta_p-\frac{\Delta_{b}}{2}-\frac{\Omega_{b}^{\text{eff}}}{2} +\frac{\Delta_{c}}{2}+\frac{\Omega_{c}^{\text{eff}}}{2})\tilde{\rho}_{C'B'} -\frac{\Omega_p}{2}\sin\theta_b\tilde{\rho}_{C'a}+\frac{\Omega_{\mu}}{2}\sin\theta_c\tilde{\rho}_{aB'}
\end{align}
\end{subequations}
In this partially diagonalized basis, we have $\tilde{\rho}_{ab}=\cos\theta_b\tilde{\rho}_{aB}-\sin\theta_b\tilde{\rho}_{aB'}$, $\tilde{\rho}_{Ca}=\cos\theta_c\tilde{\rho}_{ca}+\sin\theta_c\tilde{\rho}_{c'a}$, and $\tilde{\rho}_{C'a}=\cos\theta_c\tilde{\rho}_{c'a}-\sin\theta_c\tilde{\rho}_{ca}$.  Meanwhile,
\[
\begin{pmatrix}
\tilde{\rho}_{BB} &\tilde{\rho}_{BB'}\\
\tilde{\rho}_{B'B} &\tilde{\rho}_{B'B'}
\end{pmatrix}
=
\begin{pmatrix}
\cos^2\theta_b\tilde{\rho}_{bb}+\cos\theta_b\sin\theta_b(\tilde{\rho}_{b'b}+\tilde{\rho}_{bb'})+\sin^2\theta_b\tilde{\rho}_{b'b'} &\sin\theta_b\cos\theta_b(\tilde{\rho}_{b'b'}-\tilde{\rho}_{bb})+\cos^2\theta_b\tilde{\rho}_{bb'}-\sin^2\theta_b\tilde{\rho}_{b'b}\\
\sin\theta_b\cos\theta_b(\tilde{\rho}_{b'b'}-\tilde{\rho}_{bb})+\cos^2\theta_b\tilde{\rho}_{b'b}-\sin^2\theta_b\tilde{\rho}_{bb'} &\cos^2\theta_b\tilde{\rho}_{b'b'}-\cos\theta_b\sin\theta_b(\tilde{\rho}_{b'b}+\tilde{\rho}_{bb'})+\sin^2\theta_b\tilde{\rho}_{bb}
\end{pmatrix}.
\]
\end{widetext}

\section{}

To incorporate decay in the dressed basis, we have to make certain assumptions, as described in the text.  These amount to saying that $\gamma_{ab}\approx\gamma_{ab'}$.  and that $\gamma_b\approx\gamma_{b'}$, $\gamma^{\text{ph}}_{bc}\approx\gamma^{\text{ph}}_{b'c}$, and $\gamma^{\text{ph}}_{bc'}\approx\gamma^{\text{ph}}_{b'c'}$ so that we can take $\gamma_{cb}\approx\gamma_{cb'}=\gamma_C$ and $\gamma_{c'b}\approx\gamma_{c'b'}=\gamma_{C'}$.  Under these approximations, the contributions from decay and dephasing are
\begin{subequations}
\begin{align}
i\dot{\tilde{\rho}}_{aB}&\sim-i\gamma_{ab}\tilde{\rho}_{aB}\\
i\dot{\tilde{\rho}}_{aB'}&\sim-i\gamma_{ab}\tilde{\rho}_{aB'}\\
i\dot{\tilde{\rho}}_{CB}&\sim-i(\gamma_C\cos^2\theta_c+\gamma_{C'}\sin^2\theta_c)\tilde{\rho}_{CB}\notag\\
&\;\;-i(\gamma_{C'}-\gamma_C)\cos\theta_c\sin\theta_c\tilde{\rho}_{C'B}\\
i\dot{\tilde{\rho}}_{C'B}&\sim-i(\gamma_C\sin^2\theta_c+\gamma_{C'}\cos^2\theta_{c})\tilde{\rho}_{C'B}\notag\\
&\;\;-i(\gamma_{C'}-\gamma_C)\cos\theta_c\sin\theta_c\tilde{\rho}_{CB}\\
i\dot{\tilde{\rho}}_{CB'}&\sim-i(\gamma_C\cos^2\theta_c+\gamma_{C'}\sin^2\theta_c)\tilde{\rho}_{CB'}\notag\\
&\;\;-i(\gamma_{C'}-\gamma_C)\cos\theta_c\sin\theta_c\tilde{\rho}_{C'B'}\\
i\dot{\tilde{\rho}}_{C'B'}&\sim-i(\gamma_C\sin^2\theta_c+\gamma_{C'}\cos^2\theta_c)\tilde{\rho}_{C'B'}\notag\\
&\;\;-i(\gamma_{C'}-\gamma_C)\cos\theta_c\sin\theta_c\tilde{\rho}_{CB'}
\end{align}
\end{subequations}

\section{}

For completeness, we present here a complete and general solution for the reduced susceptibility, $\tilde{\chi}^{(1)}$, including arbitrary detunings of all fields.  We find
\begin{equation}
\tilde{\chi}^{(1)}=\frac{2\gamma_{ab}}{\Omega_p}(\cos\theta_b\tilde{\rho}_{aB}-\sin\theta_b\tilde{\rho}_{aB'}).
\end{equation}
Here,
\begin{widetext}
\begin{align}\label{fullRhoaB}
\tilde{\rho}_{aB}&=\frac{\Omega _p}{Z_+}\left(\left(\sin \theta _b  \tilde{\rho} _{B'B}-\cos\theta _b\tilde{\rho} _{BB}\right) \left(-\left(2i \gamma _{C}+\Delta _b-\Delta _c-2 \delta-\Omega_b^{\text{eff}}\right)\left(2 i\gamma _{C'}+\Delta _b-\Delta _c-2 \delta-\Omega_b^{\text{eff}}\right)\right.\right.\notag\\
&\;\;\left.\left.+2 i\cos(2 \theta _c)  \left(\gamma _{C}-\gamma _{C'}\right) \Omega_c^{\text{eff}}+(\Omega_c^{\text{eff}})^2\right)+\cos\theta _b\cos\theta _c  \tilde{\rho}_{Ca} \left(2 i\gamma_{C'}+\Delta _b- \Delta _c-2 \delta-\Omega_b^{\text{eff}}-\Omega_c^{\text{eff}}\right) \Omega _{\mu }\right.\notag\\
&\;\;\left.-\cos\theta _b  \sin \theta _c  \tilde{\rho} _{C'a} \left(2 i\gamma _{C'}+\Delta _b- \Delta _c-2\delta-\Omega_b^{\text{eff}}+\Omega_c^{\text{eff}}\right) \Omega _{\mu }\right)
\end{align}
and
\begin{align}\label{fullRhoaB'}
\tilde{\rho}_{aB'}&=\frac{\Omega _p}{Z_-} \left(\left(\sin \theta _b \tilde{\rho} _{B'B'}+\cos\theta _b\tilde{\rho}_{BB'}  \right) \left(-\left(2 i \gamma _{C}+\Delta _b-\Delta _c-2\delta+\Omega_b^{\text{eff}}\right)\left(2i \gamma _{C'}+\Delta _b-\Delta _c-\delta+\Omega_b^{\text{eff}}\right)\right.\right.\notag\\
&\;\;\left.\left.+2 i\cos(2 \theta _c)  \left(\gamma _{C}-\gamma _{C'}\right) \Omega_c^{\text{eff}}+(\Omega_c^{\text{eff}})^2\right)-\sin\theta _b\cos\theta _c  \tilde{\rho}_{Ca} \left(2i \gamma_{C'}+\Delta _b- \Delta _c-2\delta+\Omega_b^{\text{eff}}-\Omega_c^{\text{eff}}\right) \Omega _{\mu }\right.\notag\\
&\;\;\left.+\sin\theta _b  \sin \theta _c  \tilde{\rho} _{C'a} \left(2i \gamma _{C'}+\Delta _b- \Delta _c-2 \delta +\Omega_b^{\text{eff}}+\Omega_c^{\text{eff}}\right) \Omega _{\mu }\right)
\end{align}
where now
\begin{align}\label{fullZ}
Z_{\pm}&=\sin^2 \theta _c \left(2 i\gamma _{C'}+\Delta _b- \Delta _c-2 \delta\mp\Omega_b^{\text{eff}}+\Omega_c^{\text{eff}}\right) \Omega _{\mu }^2+2 i\cos^2 \theta _c \sin^2 \theta _c \left(\gamma _{C}-\gamma _{C'}\right)\left(2 i\left(\gamma _{C}-\gamma _{C'}\right) \right.\notag\\
&\;\;\times\left.\left(2 \gamma _{ab}+\Delta _b-2 \Delta _p\mp\Omega_b^{\text{eff}}\right)-\Omega _{\mu }^2\right)+\left(-2i \sin^2 \theta _c \gamma _{C}-2i \cos^2  \theta _c\gamma _{C'}-\Delta _b+ \Delta _c+2 \delta\pm\Omega_b^{\text{eff}}+\Omega_c^{\text{eff}}\right)\notag\\
&\;\;\times \left(\left(2 i\gamma_{ab}+\Delta _b-2 \Delta _p\mp\Omega_b^{\text{eff}}\right) \left(2 i\cos^2 \theta_c \gamma _{C}+2 i\sin^2 \theta _c \gamma _{C'}+\Delta _b- \Delta _c-2 \delta\mp\Omega_b^{\text{eff}}+\Omega_c^{\text{eff}}\right)-\cos^2  \theta _c\Omega _{\mu }^2\right).
\end{align}
We have defined the two photon detuning, $\delta=\Delta_p-\Delta_{\mu}=\omega_c-\omega_b+\nu_{\mu}-\nu_p$.

In the special case that $\cos\theta_b\tilde{\rho}_{BB}-\sin\theta_b\tilde{\rho}_{B'B}=\sin\theta_b\tilde{\rho}_{B'B'}+\cos\theta_b\tilde{\rho}_{BB'}$, we can define $\mathfrak{P}_B=\sqrt{2}(\cos\theta_b\tilde{\rho}_{BB}-\sin\theta_b\tilde{\rho}_{B'B})$.  Likewise, when $\sin\theta_c\tilde{\rho}_{Ca}=\cos\theta_c\tilde{\rho}_{C'a}$, we can define $\mathfrak{P}_C=-\frac{2\Omega_{\mu}}{\Omega_c^{\text{eff}}}\sin\theta_c\tilde{\rho}_{Ca}$.  Then Eqs. \ref{fullRhoaB} and \ref{fullRhoaB'} take on the relatively simple form familiar from the body of the paper,
\begin{align}\label{fullRhoaBApp}
\tilde{\rho}_{aB}&=\frac{\Omega _p}{Z_+}\left(\mathfrak{P}_{B} \left(\left(2i \gamma _{C}+\Delta _b-\Delta _c-2 \delta-\Omega_b^{\text{eff}}\right)\left(2 i\gamma _{C'}+\Delta _b-\Delta _c-2 \delta-\Omega_b^{\text{eff}}\right)-2 i\cos(2 \theta _c)  \left(\gamma _{C}-\gamma _{C'}\right) \Omega_c^{\text{eff}}\right)\right.\notag\\
&\;\;\left.+(\Omega_c^{\text{eff}})^2(\cos\theta _b\mathfrak{P}_C-\mathfrak{P}_B)\right)
\end{align}
and
\begin{align}\label{fullRhoaB'App}
\tilde{\rho}_{aB'}&=-\frac{\Omega _p}{Z_-} \left(\mathfrak{P}_{B} \left(\left(2 i \gamma _{C}+\Delta _b-\Delta _c-2\delta+\Omega_b^{\text{eff}}\right)\left(2i \gamma _{C'}+\Delta _b-\Delta _c-\delta+\Omega_b^{\text{eff}}\right)-2 i\cos(2 \theta _c)  \left(\gamma _{C}-\gamma _{C'}\right) \Omega_c^{\text{eff}}\right)\right.\notag\\
&\;\;\left.+(\Omega_c^{\text{eff}})^2(\sin\theta _b\mathfrak{P}_C-\mathfrak{P}_B)\right).
\end{align}
\end{widetext}

\bibliography{superluminal}

\end{document}